\documentclass[a4paper,11pt]{article}

\usepackage{amsmath,amssymb}     
\usepackage{color}
\usepackage{graphicx}
\usepackage{subfigure}
\usepackage{cite}                
\usepackage{hyperref}            
\usepackage{multirow,makecell}   

\numberwithin{equation}{section}   

\def \be {\begin{equation}}
\def \ee {\end{equation}}
\def \ba {\begin{array}}
\def \ea {\end{array}}
\def \bea{\begin{eqnarray}}
\def \eea{\end{eqnarray}}
\def \nn {\nonumber}

\def \a {\alpha}
\def \b {\beta}
\def \g {\gamma}
\def \c {\gamma}
\def \G {\Gamma}
\def \d {\delta}

\def \e {\epsilon}

\def \m {\mu}

\def \k {\kappa}

\def \L {\Lambda}
\def \s {\sigma}
\def \S {\Sigma}

\def \th {\theta}
\def \vth {\vartheta}

\def \t {\tau}
\def \z {\zeta}

\def \mA {\mathcal A}

\def \mD {\mathcal D}

\def \mN {\mathcal N}

\def \mP {\mathcal P}

\def \mS {\mathcal S}

\def \p {\partial}
\def \f {\frac}

\def \lt {\left}
\def \rt {\right}

\def \sr {\sqrt}
\def \td {\tilde}

\def \inf {\infty}

\def \Tr {{\textrm{Tr}}}
\def \STr {{\textrm{STr}}}
\def \diag {{\textrm{diag}}}

\def \lag {\langle}
\def \rag {\rangle}
\def \ph  {\phantom}
\def \ul  {\underline}
\def \cas {\circledast}

\def \ccc {\circledcirc}
\def \hi  {{\hat\imath}}
\def \hj  {{\hat\jmath}}

\def \Vol {\mathrm{Vol}}

\begin{document}

\title{\textbf{Supersymmetric Wilson loops in $\bf{\mN=4}$ super Chern-Simons-matter theory}}
\author{
Hao Ouyang\footnote{ouyangh@ihep.ac.cn},
Jun-Bao Wu\footnote{wujb@ihep.ac.cn}~
and
Jia-ju Zhang\footnote{jjzhang@ihep.ac.cn}
}
\date{}

\maketitle

\vspace{-8mm}

\begin{center}
{\it
 Theoretical Physics Division, Institute of High Energy Physics, Chinese Academy of Sciences,\\
19B Yuquan Rd, Beijing 100049, P.~R.~China\\\vspace{1mm}
Theoretical Physics Center for Science Facilities, Chinese Academy of Sciences,\\19B Yuquan Rd, Beijing 100049, P.~R.~China
}
\vspace{10mm}
\end{center}

\begin{abstract}

  We investigate the supersymmetric Wilson loops in $d=3$ $\mathcal{N}=4$ super Chern-Simons-matter theory obtained from non-chiral orbifold of ABJM theory. We work in both Minkowski spacetime and Euclidean space, and we construct 1/4 and 1/2 BPS Wilson loops. We also provide a complete proof that the difference between 1/4 and 1/2 Wilson loops is $Q$-exact with $Q$ being some supercharge that is preserved by both the 1/4 and 1/2 Wilson loops. This plays an important role in applying the localization techniques to compute the vacuum expectation values of Wilson loops. We also study the M-theory dual of the 1/2 BPS circular Wilson loop.

\end{abstract}

\baselineskip 18pt
\thispagestyle{empty}

\newpage

\tableofcontents

\newpage
\section{Introduction}

After the discovery of $AdS_5/CFT_4$ correspondence \cite{Maldacena:1997re,Gubser:1998bc,Witten:1998qj}, people have also been interested in the $AdS_4/CFT_3$ correspondence. A superconformal field theory (SCFT) that is dual to M-theory on $AdS_4 \times S^7$ spacetime is needed, or equivalently an SCFT that describes coinciding M2-branes is needed. The theory was finally constructed in \cite{Aharony:2008ug} and is known as Aharony-Bergman-Jafferis-Maldacena (ABJM) theory. The ABJM theory is an $\mN=6$ super Chern-Simons-matter (SCSM) theory with gauge group $U(N)\times U(N)$ and levels $(k,-k)$, and it is dual to M-theory in $AdS_4 \times S^7/Z_k$ spacetime, or type IIA superstring theory in $AdS_4 \times CP^3$ spacetime. When $k=1, 2$ the supersymmetries are enhanced non-perturbatively to $\mN=8$.

Like the study of Bogomol'nyi-Prasad-Sommerfield (BPS) Wilson loops in $d=4$ $\mN=4$ super Yang-Mills theory in $AdS_5/CFT_4$ correspondence \cite{Maldacena:1998im,Rey:1998ik,Berenstein:1998ij,Drukker:1999zq,Pestun:2007rz}, there are also many studies of Wilson loops in ABJM theory in the $AdS_4/CFT_3$ correspondence. The simplest BPS Wilson loop in ABJM theory is the 1/6 BPS one that was constructed in \cite{Drukker:2008zx,Chen:2008bp,Rey:2008bh}, but the simplest fundamental string solution dual to a Wilson loop in type IIA superstring theory in $AdS_4 \times CP^3$ spacetime is 1/2 BPS.
With more efforts the 1/2 BPS Wilson loop was constructed in \cite{Drukker:2009hy}.\footnote{There are also similar constructions of Wilson loops but with fewer supersymmetries in \cite{Griguolo:2012iq,Cardinali:2012ru,Kim:2013oza,Bianchi:2014laa,Correa:2014aga}.}
When M theory or IIA superstring theory is weakly coupled, ABJM theory is strongly coupled. Then in order to compare with available results in gravity side, one needs to compute the vacuum expectation values (VEVs) of Wilson loops at strong coupling which is usually a hard task. However, using localization techniques \cite{Kapustin:2009kz}\footnote{This has been generalized to $\mN=2$ SCSM theories in  \cite{Jafferis:2010un,Hama:2010av}.} one can compute these VEVs in ABJM theory at both weak and strong couplings \cite{Kapustin:2009kz,Marino:2009jd,Drukker:2010nc}. In applying the localization techniques to 1/2 BPS Wilson loops one needs to check that 1/2 and 1/6 BPS Wilson loops difference by a $Q$-exact term with $Q$ being some supercharge that is preserved by both the 1/2 and 1/6 BPS Wilson loops and being used in localization \cite{Drukker:2009hy}.
The result of localization techniques for the 1/2 BPS Wilson loop has also been checked up to two loops in \cite{Bianchi:2013zda,Bianchi:2013rma,Griguolo:2013sma}.

The construction of the 1/6 BPS Wilson loops in ABJM theory in \cite{Drukker:2008zx,Chen:2008bp,Rey:2008bh} is basically the same as the 1/2 BPS (1/3 BPS) Wilson loop
in $d=3$ ${\cal N}=2$ (${\cal N}=3$) Chern-Simons-matter theory in \cite{Gaiotto:2007qi}. We will call these Wilson loops Gaiotto-Yin (GY) type.
When $2\le {\cal N}\le 6$, the GY type Wilson loop usually preserves four real supercharges, including the Poncar\'e supercharges and superconformal ones,
if the loop is along a straight line or a circle\footnote{However supersymmetry enhancement for some special GY type Wilson loops in $d=3, {\cal N}=4$ theories has been recently found in \cite{Cooke:2015ila}.}.
The construction of 1/2 BPS Wilson loops in \cite{Drukker:2009hy} includes essentially the fermionic matter fields and the supergroup structure appearing in the ABJM theory. This type of Wilson loops will be called Drukker-Trancanelli (DT) type.
It is an interesting question whether DT type BPS Wilson loops exist in theories with fewer supersymmetries.
In ${\cal N}=5$ theories \cite{Aharony:2008gk,Hosomichi:2008jb}, such Wilson loops do exist and they are 2/5 BPS
\cite{Lee:2010hk}.
The situation in ${\cal N}=3$ theories is interesting but not completely clear. Based on studies of the dual M2-brane solutions, strong evidence was given in \cite{Chen:2014gta} to support the conjecture that there are no Wilson loops in ${\cal N}=3$ theories that preserve more that four supercharges.
If this conjecture is right, there are two  possibilities about the fate of the DT type BPS Wilson loops: one is that there are no DT type BPS Wilson loops in such theories, and the other is that DT type BPS Wilson loops exist in some ${\cal N}=3$ theories and they are at most 1/3 BPS. This direction is interesting to be further studied.

Putting the situations in ${\cal N}=3$ theories aside,  we would like now to study the construction of DT type BPS Wilson loops in  ${\cal N}=4$ Chern-Simons-matter theories.  Among these theories \cite{Gaiotto:2008sd,Hosomichi:2008jd}, we will focus on the theories that are obtained from the non-chiral orbifold of ABJM theory.
From ABJM theory with gauge group $U(nN)\times U(nN)$, one can perform a  $Z_n$ orbifolding  and get an $\mN=4$ SCSM theory\cite{Benna:2008zy}, and it is dual to M-theory in $AdS_4 \times S^7 /(Z_n \times Z_{nk})$ background\cite{Benna:2008zy,Imamura:2008nn,Terashima:2008ba}. The Wilson loops in this $\mN=4$ theory in fundamental representation are dual to M2-branes in $AdS_4 \times S^7 /(Z_n \times Z_{nk})$ spacetime. We consider the simplest embedding of such a membrane. The topology of the membrane worldvolume is $AdS_2\times S^1$, where $AdS_2\subset AdS_4$ and $S^1$ is along the M-theory circle direction in $S^7 /(Z_n \times Z_{nk})$. We find that there exists such an M2-brane that preserves half of the supersymmetries of the M-theory in $AdS_4 \times S^7 /(Z_n \times Z_{nk})$ spacetime. This indicates that there should be half-BPS Wilson loops in such $\mN=4$ SCSM theory. Based on experience in ABJM case, this Wilson loop may not be GY type, and may be DT type.
One of the main results in this paper is the construction of such DT type half-BPS Wilson loops.
We get the Poncar\'e and conformal supersymmetry (SUSY) transformation of the $\mN=4$ SCSM theory from that of ABJM theory. As a warm-up, we firstly construct the 1/4 BPS GY type Wilson loops. Then we give the details of the construction of the 1/2 BPS DT type Wilson loops. In Minkowski spacetime, we have the 1/4 and 1/2 BPS Wilson loops along a timelike infinite straight line. In Euclidean space, we have the 1/4 and 1/2 BPS Wilson loops along a infinite straight line, as well as
1/4 and 1/2 BPS Wilson loops along a circle. We also provide a complete proof that 1/2 and 1/4 BPS Wilson loops difference by a $Q$-exact term, with $Q$ being some supercharge that is preserved by both the 1/4 and 1/2 BPS Wilson loops.

The rest of the paper is arranged as follows. In Section~\ref{gravity}, we study the simplest M2-brane solution dual to a circular Wilson loop in orbifold ABJM theory. We compute its on-shell action with boundary terms included and we also find that this M2-brane can be half BPS. In Section~\ref{s2} we review the basics of this $\mN=4$ SCSM theory and derive its SUSY transformation. In Section~\ref{s3} we consider the 1/4 and 1/2 BPS Wilson loops along a timelike infinite straight line in Minkowski spacetime. In Section~\ref{s4} we consider the 1/4 and 1/2 BPS Wilson loops along an infinite straight line in Euclidean space. In Section~\ref{s5} we construct the circular 1/4 and 1/2 BPS Wilson loops. We conclude with conclusion and discussion in Section~\ref{s6}. We review the 1/6 and 1/2 BPS Wilson loops in ABJM theory in Appendix~\ref{abjmrev}. In Appendix~\ref{susycov} we provide a simple proof of gauge covariance of Wilson loops, from which we also prove a useful statement that has appeared in \cite{Drukker:2009hy}. In Appendix~\ref{alternative} we explore alternative definitions of Wilson loops for a super connection, but we find no nontrivial ones. In Appendix~\ref{QVappe} there are the calculation details of Subsection~\ref{QVmain}, and we give a complete proof that the difference between 1/2 and 1/4 BPS Wilson loops is $Q$-exact.

\textbf{Note added.} After the paper appears in ArXiv, there appears another paper \cite{Cooke:2015ila} that has some overlaps with ours. There are more general 1/2 BPS Wilson loops in the orbifold ABJM theory, as well as in other $\mN=4$ SCSM theories. According to terminology of \cite{Cooke:2015ila}, the 1/2 BPS Wilson loops in this paper are $\psi_1$-loops.
For a timelike straight line $x^\m=\t\d^\m_0$ in Minkowski space if we change the first two equations of the ansatz (\ref{ansatz}) to
\be
\bar\eta_i^{(2\ell)}=\bar\eta^{(2\ell)}\d^2_i, ~~~
\eta^i_{(2\ell)}=\eta_{(2\ell)}\d^i_2,
\ee
we would get the $\psi_2$-loops. In this case we would have $\hat m=\hat n=-1$, as well as
\be
\g_0 \eta_{(2\ell)}=-i\eta_{(2\ell)}, ~~~
\bar \eta^{(2\ell)} \g_0=-i\bar \eta^{(2\ell)}, ~~~
\eta_{(2\ell)} \bar \eta^{(2\ell)}=-i+\g_{0}.
\ee
The $\psi_1$- and $\psi_2$-loops have the same conserved supersymmetries (\ref{susy1}) and (\ref{susy2}).
There are similar stories for 1/2 BPS Wilson loops along straight lines and circles in Euclidean space.
There are a large number of 1/2 BPS Wilson loops, but there are not so many dual 1/2 BPS objects in M-theory. And so it is expected in \cite{Cooke:2015ila} that these Wilson loops are 1/2 BPS classically, and only some special linear combination of them is 1/2 BPS quantum mechanically.

\section{M2-branes in $AdS_4 \times S^7 /(Z_n \times Z_{nk})$ spacetime}\label{gravity}

The $\mN=4$ theory obtained from orbifolding ABJM theory is dual to M-theory in $AdS_4 \times S^7 /(Z_n \times Z_{nk})$ spacetime. We will denote $Z_n \times Z_{nk}$ as $\Gamma_{n, k}$ below. If we embed a unit $S^7$ inside ${C^4}\cong {R}^8$ as
\be \sum_{i=1}^4|z_i|^2=1, ~~~ z_i\in { C},  \ee
the action of $\G_{n, k}$ on $S^7$ is generated by \cite{Imamura:2008nn}
\be (z_1, z_2, z_3, z_4)\to (\omega_n z_1, \omega_n z_2, z_3, z_4), \ee
and \be z_i\to \omega_{nk}z_i,  \ee
where $\omega_m\equiv \exp\lt(\frac{2\pi i}{m}\rt)$.

We parameterize $z_i$ as
\bea
&& z_1 = \cos\frac{\a}{2}\cos\frac{\theta_1}{2} \exp\lt[\frac{i}4 (2\varphi_1+\chi+\zeta) \rt], \nn\\
&& z_2 = \cos\frac{\a}{2}\sin\frac{\theta_1}{2} \exp\lt[\frac{i}4 (-2\varphi_1+\chi+\zeta)\rt],\nn\\
&& z_3 = \sin\frac{\a}{2}\cos\frac{\theta_2}{2} \exp\lt[\frac{i}4 (2\varphi_2-\chi+\zeta)\rt],\\
&& z_4 = \cos\frac{\a}{2}\sin\frac{\theta_2}{2} \exp\lt[\frac{i}4 (-2\varphi_2-\chi+\zeta)\rt], \nn
\eea
where $\alpha\in [0, \pi]$, $\zeta \in [0, 8\pi]$, $\chi\in [0, 4\pi]$, $\theta_{1,2} \in [0, \pi]$, $\varphi_{1,2} \in [0, 2\pi]$.
Then the action of $\G_{n, k}$ is generated by
\be \chi\to \chi+\frac{4\pi}{n}, ~~~ \zeta\to\zeta+\frac{4\pi}{n}, \ee
and \be \zeta\to \zeta+\frac{8\pi}{nk}.\ee
The IIA limit of M-theory is obtained by taking $k\to\infty$ while keeping $n$ fixed. In this limit the circle along $\zeta$ direction will shrink.
So this circle is the M-theory circle, and its circumference is $\frac{8\pi}{nk}$.
The metric of unit $S^7$ is
\bea
&& ds^2_{S^7}=\frac14 \bigg[ d\a^2+\cos^2\frac{\a}{2} \lt( d\theta_1^2+\sin^2\theta_1 d\varphi_1^2 \rt)
                                +\sin^2\frac{\a}{2} \lt( d\theta_2^2+\sin^2\theta_2 d\varphi_2^2 \rt) \nn\\
&& \phantom{ds^2_{S^7}=\frac14 []}
+\sin^2\frac{\a}{2}\cos^2\frac{\a}{2} (d\chi+\cos\theta_1d\varphi_1-\cos\theta_2d\varphi_2)^2 \nn\\
&& \phantom{ds^2_{S^7}=\frac14 []}
+ \bigg( \frac12d\zeta+\cos^2\frac{\a}2\cos\theta_1d\varphi_1+\sin^2\frac{a}2\cos\theta_2d\varphi_2+\frac12\cos\a d\chi \bigg)^2 \bigg].
\eea
The metric of $AdS_4 \times S^7/\G_{n, k}$
is \be ds^2_{11}=R^2 \lt( \frac14 ds^2_{AdS_4}+ds^2_{S^7/\G_{n, k}} \rt).\ee
For Lorentzian signature we choose the following global coordinate on $AdS_4$,
\be \label{mads4}
 ds^2_{AdS_4}=\cosh^2u \lt(-\cosh^2\rho dt^2+d\rho^2 \rt)+du^2+\sinh^2u d\phi^2.
\ee
The four-form field strength on this background is
\be H_4=\frac{3R^3}8 \cosh^2u\sinh u\cosh\rho dt\wedge d\rho \wedge du \wedge d\phi . \ee
Flux quantization gives
\be
R = 2\pi \ell_p \left[ \frac{N}{6 \Vol(S^7/\G_{n, k})}\right]^{1/6}
  = \ell_p (32 \pi^2 n^2 Nk)^{1/6}, \label{rlp}
\ee
where $\ell_p$ is the eleven-dimensional Planck length, and we have used
\be \Vol(S^7/\G_{n, k})=\frac{\Vol(S^7)}{n^2k}=\frac{\pi^4}{3n^2k}. \ee
The radius of the $\zeta$ circle in Planck unit is of order
$R/(nk\ell_p) \propto (n^2 Nk)^{1/6}/(nk)$, and so the M-theory description is a
good one when $N \gg n^4k^5$.

We consider the probe M2-brane solution in this background.
In Lorentzian signature the bosonic part of the M2-brane action is
\be
S_{M2}=S_{M2}^{DBI}+S_{M2}^{WZ}=- T_{M2}\left(\int d^3\s\sqrt{-\mbox{det}g_{mn}} + \int P[C_3]\right).
\ee
Here $g_{mn}$ is the induced metric of the membrane worldvolume, $T_{M2}$ is
the tension of the M2-brane
\be T_{M2}={1\over (2\pi)^2\ell_p^3}, \ee
and $P[C_3]$ is the pullback of the bulk 3-form gauge
potential to the worldvolume of the membrane.
The gauge choice for the background 3-form gauge potential  $C_3$ is
\be C_3=\frac{R^3}8(\cosh^3u-1)\cosh\rho dt\wedge d\rho\wedge d\phi.
\ee
From the action, one can obtain the
membrane equation of motion as\footnote{We always use the indices from the beginning (middle) of the alphabet to refer
to the frame (coordinate) coordinates, and the underlined indices to
refer to the target space ones.}
\be
\frac{1}{\sqrt{-g}}\p_m\left(\sqrt{-g}g^{mn}\p_nX^{\underline{N}}\right)G_{\underline{MN}}
+g^{mn}\p_{m}X^{\underline{N}}\p_{n}X^{\underline{P}}\G^{\underline{Q}}_{\underline{NP}}G_{\underline{QM}}
=\frac{1}{3!\sqrt{-g}}\epsilon^{mnp}(P[d C_3])_{\underline{M}mnp}.
\ee
Note that $\epsilon^{mnp}$ is a tensor density on the world-volume of the membrane.

Since we want to find the simplest membrane embedding corresponding to
a Wilson loop in the  dual field theory, we take the topology of the membrane worldvolume to be $AdS_2\times S^1$. The $AdS_2$ is embedded in
$AdS_4$, while $S^1$ is along the M-theory circle.
So we consider the ansatz
\be t=\sigma^0, ~~~ \rho=\sigma^1, ~~~ \zeta=\sigma^2,  \ee
where $\sigma^\mu, \mu=0, 1, 2$ is the coordinates on the worldvolume of M2-brane.
One can find that the equations of motion only lead to the constraint that $u=0$.
Then the induced metric of M2-brane is
\be d s_{M2}^2=R^2 \lt(-\frac14\cosh^2\rho dt^2+\frac14d\rho^2+\frac1{16}d\zeta^2 \rt).  \ee

To compute the on-shell action of the M2-brane whose boundary at infinity is $S^1$, we work in Euclidean signature and choose the $AdS_4$ coordinates
\be ds^2_{AdS_4}=\cosh^2u \lt(\sinh^2\rho d\psi^2+d\rho^2 \rt)+du^2+\sinh^2u d\phi^2, \ee
where $\psi \in [0,2\pi]$.
In Euclidean signature the M2-brane action becomes\footnote{It is easy to see that $S_{M2}^{WZ}=0$ for the M2-brane solution considered here.}
\be
S_{M2}=S_{M2}^{DBI}=T_2 \int d^3\s\sqrt{\mbox{det}g_{mn}}.
\ee
For the M2-brane that is put at
\be \psi=\sigma_1, ~~~ \rho=\sigma_2, ~~~ \zeta=\sigma_3, ~~~ u=0,  \ee
the on-shell action is
\be S_{M2}=\frac{T_{M2}R^3}{16}\int d\zeta d\rho d\psi\sinh \rho.\ee
After adding boundary terms to regulate the action as in \cite{Drukker:1999zq}, we get
\bea S^{total}_{M2}=-\frac{\pi T_{M2}R^3}{8}\int d\zeta.\eea
Using the fact that $\zeta \in[0, \frac{8\pi}{nk}]$, $T_{M2}=1/(4\pi^2\ell_p^3)$ and (\ref{rlp}), we can get
\be  S^{total}_{M2}=-\pi\sqrt{\frac{2N}{k}}.\ee
Then the holographic prediction for the leading exponential behavior of the VEV of the 1/2 BPS Wilson loop in the large $N$ limit with finite $k$ and $n$ is
\be \label{m2exp}
\lag W \rag \sim \exp(-S^{total}_{M2})= \exp(\pi\sqrt{2N/k}).
\ee
Note that the result is not dependent on $n$.

In Lorentzian signature the Killing spinor of M-theory in $AdS_4\times S^7$ spacetime with the $AdS_4$ coordinates (\ref{mads4}) is \cite{Drukker:2008zx}
\be
\epsilon = e^{\frac{\a}{4}(\hat{\gamma}\g_4-\g_{7\sharp})}
           e^{\frac{\theta_1}4 (\hat{\g}\g_5-\g_{8\sharp})}
           e^{\frac{\theta_2}2(\g_{79}+\g_{46}) }
           e^{-\frac{\xi_1}2\hat{\g}\g_{\sharp}}e^{-\frac{\xi_2}{2}\g_{58}}
           e^{-\frac{\xi_3}2\g_{47}}e^{-\frac{\xi_4}2\g_{69}}
           e^{\frac{u}2\hat{\g}\g_2}e^{\frac{\phi}2\g_{23}}
           e^{\frac{\rho}2\hat{\g}\g_{1}}e^{\frac{t}2\hat{\g}\g_0}
           \epsilon_0,
\ee
where $\epsilon_0$ is a constant eleven-dimensional Majorana spinor which has 32 real degrees of freedom. The definitions of $\xi_i$ with $i=1,2,3,4$ are
\bea
&&  \xi_1 = \frac14(2\phi_1+\chi+\zeta),  ~~~
    \xi_2 = \frac14(-2\phi_1+\chi+\zeta), \nn\\
&&  \xi_3 = \frac14(2\phi_2-\chi+\zeta),  ~~~
    \xi_4 = \frac14(-2\phi_2-\chi+\zeta).
\eea
Also $\g_0$, $\g_1$, $\cdots$, $\g_9$, $\g_\sharp$ are eleven-dimensional gamma matrices, and $\hat{\gamma}\equiv \g^{0123}$. Note that the eleven-dimensional gamma matrices are chosen such that
\be \g_{0123456789\sharp}=1. \ee

To obtained the Killing spinor of M-theory on  $AdS_4\times S^7/\G_{n, k}$, we need to impose the conditions
\be \label{e11}
{\cal L}_{K_1}\epsilon={\cal L}_{K_2}\epsilon=0,
\ee
where $K_{1,2}$ are the following two Killing vectors
\be
K_1 = \p_{\chi}+\p_{\z}, ~~~
K_2 = \p_\zeta,
\ee
and they are related to the generators of $\G_{n, k}$.
The definition of ${\cal L}_{K}\epsilon $ is
\be {\cal L}_{K}\epsilon\equiv K^{\ul{M}}\nabla_{\ul{M}}\epsilon+\frac14(\nabla_{\ul{M}}K_{\ul{N}})\g^{\ul{M}\ul{N}}\epsilon.\ee
After some computations, we find that the two conditions (\ref{e11}) are equivalent to
\be
\g_{4679} \epsilon_0 = -\epsilon_0. \label{proj1}
\ee
So the background is half BPS compared to the maximal possibility, i.e.\ there are 16 real supercharges.
This is consistent with the fact that the dual three-dimensional SCFT is an $\mN=4$ theory.

 The supercharges preserved by the probe membrane are determined by the following equation
\be \Gamma_{M2}\epsilon=\epsilon,\ee
with \be\Gamma_{M2}= \frac{1}{\sqrt{-g}}\partial_{\sigma_0} X^{\ul{M}}\partial_{\sigma_1} X^{\ul{N}}\partial_{\sigma_2} X^{\ul{P}}e^{\ul{A}}_{\ul{M}}e^{\ul{B}}_{\ul{N}}e^{\ul{C}}_{\ul{P}}\g_{\ul{A}\ul{B}\ul{C}}. \ee
For the membrane we just found, we have \be\G_{M2}=\g_{01\sharp}. \ee
So the supercharges preserved by this probe membrane  correspond to the solution of \be \g_{01\sharp}\epsilon=\epsilon. \ee
At the positions with $\alpha=\theta_1=0$, this is equivalent to \cite{Drukker:2008zx}
\be \g_{01\sharp}\epsilon_0=\epsilon_0. \ee
Since it is  compatible with the projection condition in (\ref{proj1}), we arrive at the conclusion that the probe M2-brane put at $\a=\theta_1=0$ is half BPS compared to the supersymmetries of M-theory in $AdS_4\times S^7/\G_{n, k}$ spacetime.

\section{$\mN=4$ SCSM theory}\label{s2}

Orbifolding the ABJM theory with gauge group  $U(nN)\times U(nN)$  and levels $(k,-k)$ by $Z_n$,
one can get the $\mN=4$ SCSM theory with gauge group $U(N)^{2n}$ and Chern-Simons levels $(k, -k, \cdots, k, -k)$ \cite{Benna:2008zy}.
We can get the SUSY transformation of this $\mN=4$ theory from that of ABJM theory (\ref{abjmsusy}) by the orbifolding. The result is
\bea \label{ne4susy}
&& \d\phi_i^{(2\ell+1)}=2i\bar\chi_{i\hi}\psi^\hi_{(2\ell+1)}, ~~~
   \d\phi_\hi^{(2\ell)}=-2i\bar\chi_{i\hi}\psi^i_{(2\ell)}, \nn\\
&& \d\bar\phi^i_{(2\ell+1)}=2i\bar\psi_\hi^{(2\ell+1)}\chi^{i\hi}, ~~~
   \d\bar\phi^\hi_{(2\ell)}=-2i\bar\psi_i^{(2\ell)}\chi^{i\hi},  \nn\\
&& \d A_\m^{(2\ell+1)}=\f{4\pi}{k} \lt[ \lt( \phi_i^{(2\ell+1)}\bar\psi_\hi^{(2\ell+1)}
                                            -\phi_\hi^{(2\ell)}\bar\psi_i^{(2\ell)} \rt) \g_\m \chi^{i\hi}
                                        +\bar\chi_{i\hi}\g_\m \lt( \psi^\hi_{(2\ell+1)}\bar\phi^i_{(2\ell+1)}
                                                         -\psi^i_{(2\ell)}\bar\phi^\hi_{(2\ell)} \rt)\rt],    \nn\\
&& \d\hat A_\m^{(2\ell)}=\f{4\pi}{k} \lt[ \lt( \bar\psi_\hi^{(2\ell-1)}\phi_i^{(2\ell-1)}
                                              -\bar\psi_i^{(2\ell)}\phi_\hi^{(2\ell)} \rt) \g_\m\chi^{i\hi}
+\bar\chi_{i\hi}\g_\m \lt( \bar\phi^i_{(2\ell-1)}\psi^\hi_{(2\ell-1)}-\bar\phi^\hi_{(2\ell)}\psi^i_{(2\ell)} \rt) \rt], \nn\\
&& \d\psi^i_{(2\ell)}=2\g^\m\chi^{i\hi}D_\m\phi_\hi^{(2\ell)} +2 \vth^{i\hi}\phi_\hi^{(2\ell)}
            -\f{4\pi}{k}\chi^{i\hi} \lt(  \phi_\hi^{(2\ell)}\bar\phi^j_{(2\ell-1)}\phi_j^{(2\ell-1)} \rt.   \nn\\
&& \phantom{\d\psi^i_{(2\ell)}=}   \lt. +\phi_\hi^{(2\ell)}\bar\phi^\hj_{(2\ell)}\phi_\hj^{(2\ell)}
                                        -\phi_j^{(2\ell+1)}\bar\phi^j_{(2\ell+1)}\phi_\hi^{(2\ell)}
                                        -\phi_\hj^{(2\ell)}\bar\phi^\hj_{(2\ell)}\phi_\hi^{(2\ell)}  \rt)   \nn\\
&& \phantom{\d\psi^i_{(2\ell)}=}
            -\f{8\pi}{k}\chi^{j\hj} \lt(  \phi_j^{(2\ell+1)}\bar\phi^i_{(2\ell+1)}\phi_\hj^{(2\ell)}
                                        -\phi_\hj^{(2\ell)}\bar\phi^i_{(2\ell-1)}\phi_j^{(2\ell-1)} \rt),   \nn\\
&& \d\psi^\hi_{(2\ell+1)}= -2\g^\m\chi^{i\hi}D_\m\phi_i^{(2\ell+1)} -2\vth^{i\hi}\phi_i^{(2\ell+1)}
+\f{4\pi}{k}\chi^{i\hi} \lt( \phi_i^{(2\ell+1)}\bar\phi^j_{(2\ell+1)}\phi_j^{(2\ell+1)} \rt.                \nn\\
&& \phantom{\d\psi^\hi_{(2\ell+1)}=} \lt.  +\phi_i^{(2\ell+1)}\bar\phi^\hj_{(2\ell+2)}\phi_\hj^{(2\ell+2)}
                                           -\phi_j^{(2\ell+1)}\bar\phi^j_{(2\ell+1)}\phi_i^{(2\ell+1)}
                                           -\phi_\hj^{(2\ell)}\bar\phi^\hj_{(2\ell)}\phi_i^{(2\ell+1)} \rt) \nn\\
&& \phantom{\d\psi^\hi_{(2\ell+1)}=}
            -\f{8\pi}{k}\chi^{j\hj} \lt(  \phi_j^{(2\ell+1)}\bar\phi^\hi_{(2\ell+2)}\phi_\hj^{(2\ell+2)}
                                         -\phi_\hj^{(2\ell)}\bar\phi^\hi_{(2\ell)}\phi_j^{(2\ell+1)} \rt),   \\
&& \d\bar\psi_i^{(2\ell)}=-2\bar\chi_{i\hi}\g^\m D_\m\bar\phi^\hi_{(2\ell)} +2\bar\vth_{i\hi}\bar\phi^\hi_{(2\ell)}
+\f{4\pi}{k}\bar\chi_{i\hi} \lt(  \bar\phi^\hi_{(2\ell)}\phi_j^{(2\ell+1)}\bar\phi^j_{(2\ell+1)}   \rt.            \nn\\
&&\phantom{\d\psi_i^{(2\ell)}=}    \lt.  +\bar\phi^\hi_{(2\ell)}\phi_\hj^{(2\ell)}\bar\phi^\hj_{(2\ell)}
                                         -\bar\phi^j_{(2\ell-1)}\phi_j^{(2\ell-1)}\bar\phi^\hi_{(2\ell)}
                                         -\bar\phi^\hj_{(2\ell)}\phi_\hj^{(2\ell)}\bar\phi^\hi_{(2\ell)} \rt)      \nn\\
&& \phantom{\d\psi_i^{(2\ell)}=}
            +\f{8\pi}{k}\bar\chi_{j\hj} \lt(  \bar\phi^j_{(2\ell-1)}\phi_i^{(2\ell-1)}\bar\phi^\hj_{(2\ell)}
                                            -\bar\phi^\hj_{(2\ell)}\phi_i^{(2\ell+1)}\bar\phi^j_{(2\ell+1)} \rt),  \nn\\
&& \d\bar\psi_\hi^{(2\ell+1)}= 2\bar\chi_{i\hi}\g^\m D_\m\bar\phi^i_{(2\ell+1)} -2\bar\vth_{i\hi}\bar\phi^i_{(2\ell+1)}
         -\f{4\pi}{k}\bar\chi_{i\hi} \lt(  \bar\phi^i_{(2\ell+1)}\phi_j^{(2\ell+1)}\bar\phi^j_{(2\ell+1)}  \rt.  \nn\\
&& \phantom{\d\bar\psi_\hi^{(2\ell+1)}=} \lt. +\bar\phi^i_{(2\ell+1)}\phi_\hj^{(2\ell)}\bar\phi^\hj_{(2\ell)}
                                              -\bar\phi^j_{(2\ell+1)}\phi_j^{(2\ell+1)}\bar\phi^i_{(2\ell+1)}
                                              -\bar\phi^\hj_{(2\ell+2)}\phi_\hj^{(2\ell+2)}\bar\phi^i_{(2\ell+1)} \rt) \nn\\
&& \phantom{\d\bar\psi_\hi^{(2\ell+1)}=}
            +\f{8\pi}{k}\bar\chi_{j\hj} \lt(  \bar\phi^j_{(2\ell+1)}\phi_\hi^{(2\ell)}\bar\phi^\hj_{(2\ell)}
                                            -\bar\phi^\hj_{(2\ell+2)}\phi_\hi^{(2\ell+2)}\bar\phi^j_{(2\ell+1)} \rt). \nn
\eea
Here $\ell=0,1, \cdots, n-1 $. There are no summations of $\ell$ here, and would not be summations of $\ell$ later unless it is given out explicitly.
Indices $i,j,\cdots=1,2$ and $\hi,\hj,\cdots=\hat 1,\hat 2$ are those of the $SU(2)\times SU(2)$ R-symmetry.
The definitions of covariant derivatives are
\bea
&& D_\m \phi_\hi^{(2\ell)} =\p_\m \phi_\hi^{(2\ell)} +i A_\m^{(2\ell+1)} \phi_\hi^{(2\ell)}
                                                    -i \phi_\hi^{(2\ell)} \hat A_\m ^{(2\ell)} ,\nn\\
&& D_\m \phi_i^{(2\ell+1)} =\p_\m \phi_i^{(2\ell+1)} +i A_\m^{(2\ell+1)} \phi_i^{(2\ell+1)}
                                                    -i \phi_i^{(2\ell+1)} \hat A_\m ^{(2\ell+2)} ,\nn\\
&& D_\m \bar\phi^\hi_{(2\ell)} =\p_\m \bar\phi^\hi_{(2\ell)} +i \hat A_\m ^{(2\ell)} \bar\phi^\hi_{(2\ell)}
                                                             -i \bar\phi^\hi_{(2\ell)}A_\m^{(2\ell+1)} ,\\
&& D_\m \bar\phi^i_{(2\ell+1)} =\p_\m \bar\phi^i_{(2\ell+1)} +i \hat A_\m ^{(2\ell+2)} \bar\phi^i_{(2\ell+1)}
                                                             -i \bar\phi^i_{(2\ell+1)}A_\m^{(2\ell+1)}. \nn
\eea
Also $\chi^{i\hi}=\th^{i\hi}+x^\m\g_\m\vth^{i\hi}$ and $\bar\chi_{i\hi}=\bar\th_{i\hi}-\bar \vth_{i\hi}x^\m\g_\m$, and $\th^{i\hi}$, $\bar\th_{i\hi}$, $\vth^{i\hi}$, $\bar\vth_{i\hi}$ are Dirac spinors with constraints
\bea \label{thihi}
&& (\th^{i\hi})^*=\bar \th_{i\hi}, ~~~ \bar\th_{i\hi}=\e_{ij}\e_{\hi\hj}\th^{j\hj},  \nn\\\
&& (\vth^{i\hi})^*=\bar \vth_{i\hi}, ~~~ \bar\vth_{i\hi}=\e_{ij}\e_{\hi\hj}\vth^{j\hj}.
\eea
Symbols $\e_{ij}$ and $\e_{\hi\hj}$ are antisymmetric with $\e_{12}=\e_{\hat 1 \hat 2}=1$.
Note that for $\d$ in (\ref{ne4susy}) we have
\be \label{j14}
\d=2i \lt( \bar \th_{i\hi} P^{i\hi} + \bar \vth_{i\hi} S^{i\hi} \rt)
  =2i \lt( \bar P_{i\hi} \th^{i\hi} + \bar S_{i\hi} \vth^{i\hi} \rt),
\ee
with $P^{i\hi}$, $\bar P_{i\hi}$ and $S^{i\hi}$, $S_{i\hi}$ being Poncar\'e and conformal supercharges that satisfy
\bea \label{j15}
&& (P^{i\hi})^*=\bar P_{i\hi}, ~~~ \bar P_{i\hi}=\e_{ij}\e_{\hi\hj} P^{j\hj},  \nn\\\
&& (S^{i\hi})^*=\bar S_{i\hi}, ~~~ \bar S_{i\hi}=\e_{ij}\e_{\hi\hj} S^{j\hj}.
\eea

In Euclidean space, the SUSY transformation is formally identical to (\ref{ne4susy}),
with $\chi^{i\hi}=\th^{i\hi}+x^\m\g_\m\vth^{i\hi}$ and $\bar\chi_{i\hi}=\bar\th_{i\hi}-\bar \vth_{i\hi}x^\m\g_\m$.
But now equations (\ref{thihi}) become
\be
\bar\th_{i\hi}=\e_{ij}\e_{\hi\hj}\th^{j\hj}, ~~~ \nn\\\
\bar\vth_{i\hi}=\e_{ij}\e_{\hi\hj}\vth^{j\hj}.
\ee
Note the eight spinors $\th^{i\hi}$, $\vth^{i\hi}$ with $i=1,2$, $\hi=\hat 1, \hat 2$ are independent Dirac spinors.
Now equations (\ref{j14}) are invariant but (\ref{j15}) become
\be
\bar P_{i\hi}=\e_{ij}\e_{\hi\hj} P^{j\hj},  ~~~
\bar S_{i\hi}=\e_{ij}\e_{\hi\hj} S^{j\hj}.
\ee

\section{Straight line in Minkowski spacetime}\label{s3}

In Minkowski spacetime there are BPS Wilson loops along null and timelike infinite straight lines \cite{Ouyang:2015ada}. It is easy to construct a null Wilson loop, and it is 1/2 BPS. We only consider the timelike BPS Wilson loops here.

\subsection{1/4 BPS Wilson loop}

We consider the Wilson loop along a timelike straight line $x^\m=\t\d^\m_0$ as
\bea  \label{w14a}
&& W_{1/4}^{(2\ell+1)}=\mP \exp \lt( -i\int d\t \mA^{(2\ell+1)}(\t)  \rt),\\
&& \mA^{(2\ell+1)}=A_\m^{(2\ell+1)}\dot x^\m +\f{2\pi}{k} \lt(
                                        M^i_{\ph{i}j} \phi_i^{(2\ell+1)}\bar\phi^j_{(2\ell+1)}
                                       +M^\hi_{\ph{\hi}\hj} \phi_\hi^{(2\ell)}\bar\phi^\hj_{(2\ell)}  \rt) |\dot x|. \nn
\eea
For Poncar\'e SUSY transformation we can get
\bea
&& \d\mA^{(2\ell+1)}=\f{4\pi}{k} \lt[  \phi_i^{(2\ell+1)}\bar\psi_\hi^{(2\ell+1)} \lt(  \g_0\th^{i\hi}+i M^i_{\ph ij}\th^{j\hi} \rt)
                                      -\phi_\hi^{(2\ell)}\bar\psi_i^{(2\ell)} \lt(  \g_0\th^{i\hi}+i M^\hi_{\ph \hi\hj}\th^{i\hj} \rt) \rt. \nn\\
&&\phantom{\d\mA^{(2\ell+1)}=\f{4\pi}{k} \lt[\rt.} \lt.
                                   +\lt( \bar\th_{i\hi}\g_0+iM^j_{\ph j i}\bar\th_{j\hi} \rt)\psi^\hi_{(2\ell+1)}\bar\phi^i_{(2\ell+1)}
                                   -\lt( \bar\th_{i\hi}\g_0+iM^\hj_{\ph \hj \hi}\bar\th_{i\hj} \rt)\psi^i_{(2\ell)}\bar\phi^\hi_{(2\ell)} \rt].
\eea
We work in the basis of diagonal $M^i_{\ph ij}= m_i \d^i_j$ and $M^\hi_{\ph \hi \hj}= m_\hi \d^\hi_\hj$, and then we get
\bea \label{e7}
&& \g_0 \th^{i\hi}=-i m_i \th^{i\hi}=-i m_\hi \th^{i\hi}, \nn\\
&& \g_0 \bar\th_{i\hi}=i  m_i \bar\th_{i\hi}=i  m_\hi \bar\th_{i\hi}.
\eea
Supposing $\th^{1\hat 1}\neq 0$, we choose without loss of generality
\be
\g_0\th^{1\hat 1}=i\th^{1\hat 1}.
\ee
Using (\ref{thihi}) we know $(\th^{1\hat1})^*=\th^{2\hat2}$, and then we get
\be
\g_0\th^{2\hat 2}=-i\th^{2\hat 2}.
\ee
This means that $ m_1= m_{\hat 1}=-1$, $ m_2= m_{\hat 2}=1$. Then we have
\be
\th^{1\hat 2}=\th^{2\hat 1}=0.
\ee
We can check that the equations (\ref{e7}) are consistent.
It is similar for conformal SUSY transformation.
Thus we get a 1/4 BPS Wilson loop.

Similarly we can construct the 1/4 BPS Wilson loop along $x^\m= \t\d^\m_0$ that preserves the same supersymmetries
\bea \label{w14ha}
&& \hat W_{1/4}^{(2\ell)}=\mP \exp \lt( -i\int d\t \hat\mA^{(2\ell)} (\t) \rt),\nn\\
&& \hat\mA^{(2\ell)}=\hat A_\m^{(2\ell)}\dot x^\m +\f{2\pi}{k} \lt(
                                            N_i^{\ph{i}j}\bar\phi^i_{(2\ell-1)}\phi_j^{(2\ell-1)}
                                           +N_\hi^{\ph{\hi}\hj}\bar\phi^\hi_{(2\ell)}\phi_\hj^{(2\ell)}  \rt) |\dot x|, \\
&& N_i^{\ph ij}=N_\hi^{\ph \hi\hj}=\diag ( -1, 1 ). \nn
\eea

Also we can combine (\ref{w14a}) and (\ref{w14ha}) and get the 1/4 BPS Wilson loop
\bea \label{w14ne4}
&& W_{1/4}=\mP \exp \lt( -i\int d\t L_{1/4}(\t) \rt),\nn\\
&& L_{1/4}=\lt( \ba{cc} \mA & \\ & \hat\mA \ea \rt), \\
&& \mA=\diag (\mA^{(1)},\mA^{(3)},\cdots,\mA^{(2n-1)}), \nn\\
&& \hat\mA=\diag (\hat\mA^{(0)},\hat\mA^{(2)},\cdots,\hat\mA^{(2n-2)}). \nn
\eea
Note that we can also construct the 1/4 BPS Wilson loop in this subsection using the consideration in \cite{Gaiotto:2007qi}
for general $\mN=2$ theories. So this Wilson loop is GY type.

\subsection{1/2 BPS Wilson loop}

We consider the timelike Wilson loop along $x^\m=\t\d^\m_0$
\be \label{w12ne4}
W_{1/2}=\mP \exp \lt( -i\int d\t L_{1/2}(\t) \rt),
\ee
where $L_{1/2}$ is a supermatrix
\be
L_{1/2}=\lt( \ba{cc} \mA & \bar F_1 \\ F_2 & \hat\mA \ea \rt).
\ee
Here we have definitions
\bea \label{eqlargea}
&& \mA=\diag (\mA^{(1)},\mA^{(3)},\cdots,\mA^{(2n-1)}), \nn\\
&& \mA^{(2\ell+1)}=A_\m^{(2\ell+1)} \dot x^\m +\f{2\pi}{k} \lt(  M^i_{\ph{i}j} \phi_i^{(2\ell+1)}\bar\phi^j_{(2\ell+1)}
                                              +M^\hi_{\ph{\hi}\hj} \phi_\hi^{(2\ell)}\bar\phi^\hj_{(2\ell)}  \rt)|\dot x|, \nn\\
&& \hat\mA= \diag (\hat\mA^{(0)},\hat\mA^{(2)},\cdots,\hat\mA^{(2n-2)}) ,  \\
&& \hat\mA^{(2\ell)}=\hat A_\m^{(2\ell)}\dot x^\m +\f{2\pi}{k} \lt(  N_i^{\ph{i}j}\bar\phi^i_{(2\ell-1)}\phi_j^{(2\ell-1)}
                                        +N_\hi^{\ph{\hi}\hj}\bar\phi^\hi_{(2\ell)}\phi_\hj^{(2\ell)}  \rt)|\dot x|, \nn
\eea
as well as
\bea
&& \bar F_1=\lt( \ba{ccccc}
\bar f_1^{(0)}&\bar f_1^{(1)}&&& \\
&\bar f_1^{(2)}&\bar f_1^{(3)}&& \\
&&\ddots&\ddots& \\
&&&\bar f_1^{(2n-4)}&\bar f_1^{(2n-3)} \\
\bar f_1^{(2n-1)}&&&&\bar f_1^{(2n-2)}
\ea \rt) |\dot x|,                                          \nn\\
&& \bar f_1^{(2\ell+1)}=\sr{\f{2\pi}{k}}\bar\eta_\hi^{(2\ell+1)}\psi^\hi_{(2\ell+1)}, ~~~
   \bar f_1^{(2\ell)}=\sr{\f{2\pi}{k}}\bar\eta_i^{(2\ell)}\psi^i_{(2\ell)},                   \nn\\
&& F_2=\lt( \ba{ccccc}
f_2^{(0)}&&&&f_2^{(2n-1)} \\
f_2^{(1)}&f_2^{(2)}&&& \\
&f_2^{(3)}&\ddots&& \\
&&\ddots&f_2^{(2n-4)}& \\
&&&f_2^{(2n-3)}&f_2^{(2n-2)}
\ea \rt) |\dot x|,                                          \\
&& f_2^{(2\ell+1)}=\sr{\f{2\pi}{k}}\bar\psi_\hi^{(2\ell+1)}\eta^\hi_{(2\ell+1)}, ~~~
   f_2^{(2\ell)}=\sr{\f{2\pi}{k}}\bar\psi_i^{(2\ell)}\eta^i_{(2\ell)}.                        \nn
\eea
Note that $\mA$ and $\hat\mA$ are Grassmann even. Also $\bar\eta_\hi^{(2\ell+1)}$, $\bar\eta_i^{(2\ell)}$, $\eta^\hi_{(2\ell+1)}$ and  $\eta^i_{(2\ell)}$ are Grassmann even, and so $\bar F_1$ and $F_2$ are Grassmann odd.
To make $W_{1/2}$ SUSY invariant, we need\cite{Lee:2010hk}
\be
\d L_{1/2}= \mD_\t G \equiv \p_\t G+i[L_{1/2},G],
\ee
for some Grassmann odd supermatrix
\be
G= \lt( \ba{cc} & \bar G_1 \\ G_2 &  \ea \rt).
\ee
Concretely, we need
\bea \label{e9}
&& \d \mA=i(\bar F_1 G_2-\bar G_1 F_2),  \nn\\
&& \d \hat\mA=i(F_2 \bar G_1 - G_2 \bar F_1),  \\
&& \d \bar F_1=\mD_\t \bar G_1 \equiv \p_\t \bar G_1+i\mA \bar G_1-i\bar G_1\hat\mA,  \nn\\
&& \d F_2=\mD_\t G_2 \equiv \p_\t G_2+i\hat\mA G_2-i G_2 \mA.  \nn
\eea

As in \cite{Drukker:2009hy}, we can use symmetry to guide the search for a 1/2 BPS Wilson loop.
We break the $SU(2)\times SU(2)$ R-symmetry to $U(1)\times SU(2)$ by writing $(i,\hi)=(1,2,\hi)$.
We wish to get a BPS Wilson loop with the $SU(2)$ subgroup intact, and so we choose
\bea \label{ansatz}
&& \bar\eta_i^{(2\ell)}=\bar\eta^{(2\ell)}\d^1_i, ~~~
   \eta^i_{(2\ell)}=\eta_{(2\ell)}\d^i_1, ~~~
   \bar\eta_\hi^{(2\ell+1)}=\eta^\hi_{(2\ell+1)}=0, \nn\\
&& M^i_{\ph i j}=\diag( m_1, m_2 ), ~~~
   M^\hi_{\ph \hi \hj}=\diag( \hat m, \hat m ),  \\
&& N_i^{\ph{i}j}=\diag( n_1, n_2 ), ~~~
   N_\hi^{\ph{\hi}\hj}=\diag( \hat n, \hat n ).  \nn
\eea
Then we need
\bea
&& \bar G_1= \diag (\bar g_1^{(0)},\bar g_1^{(2)}, \cdots, \bar g_1^{(2n-2)} ),   \nn\\
&& G_2= \diag (g_2^{(0)},g_2^{(2)}, \cdots, g_2^{(2n-2)} ).
\eea
And then equations (\ref{e9}) become
\bea \label{e10}
&& \d \mA^{(2\ell+1)}=i(\bar f_1^{(2\ell)} g_2^{(2\ell)}-\bar g_1^{(2\ell)}f_2^{(2\ell)}),  \nn\\
&& \d \hat\mA^{(2\ell)}=i(f_2^{(2\ell)} \bar g_1^{(2\ell)} - g_2^{(2\ell)} \bar f_1^{(2\ell)}),\\
&& \d \bar f_1^{(2\ell)}=\mD_\t \bar g_1^{(2\ell)}
             \equiv \p_\t \bar g_1^{(2\ell)}+i\mA^{(2\ell+1)} \bar g_1^{(2\ell)}-i\bar g_1^{(2\ell)}\hat\mA^{(2\ell)},  \nn\\
&& \d f_2^{(2\ell)}=\mD_\t g_2^{(2\ell)}
             \equiv \p_\t g_2^{(2\ell)}+i\hat\mA^{(2\ell)} g_2^{(2\ell)}-ig_2^{(2\ell)}\mA^{(2\ell+1)}.\nn
\eea

Without loss of generality, we suppose that
\be \label{susy1}
\g_0\th^{1\hi}=i\th^{1\hi}, ~~ \hi=\hat1,\hat2,
\ee
and then from (\ref{thihi}) we have
\be \label{susy2}
\g_0\th^{2\hi}=-i\th^{2\hi}, ~~~\bar \th_{1\hi}\g_0=i\bar\th_{1\hi}, ~~~ \bar \th_{2\hi}\g_0=-i\bar\th_{2\hi}.
\ee
For $\psi^2$, $\psi^\hi$ and $\bar\psi_2$, $\bar\psi_\hi$ not appearing in $\d\mA^{(2\ell+1)}$ and $\d\hat\mA^{(2\ell)}$, we have to choose $ m_1=n_1=-1$ and $m_2=\hat m= n_2=\hat n=1$, and then we get
\bea
&& \d\mA^{(2\ell+1)}=-\f{8\pi i}{k} \lt( \phi_\hi^{(2\ell)}\bar\psi_1^{(2\ell)}\th^{1\hi}
                                         +\bar\th_{1\hi}\psi^1_{(2\ell)}\bar\phi^\hi_{(2\ell)} \rt), \nn\\
&& \d\hat\mA^{(2\ell)}=-\f{8\pi i}{k} \lt(  \bar\psi_1^{(2\ell)}\phi_\hi^{(2\ell)}\th^{1\hi}
                                           +\bar\th_{1\hi}\bar\phi^\hi_{(2\ell)}\psi^1_{(2\ell)}  \rt).
\eea
For $\d \bar f_1^{(2\ell)}$  and $\d f_2^{(2\ell)}$ satisfying the form of (\ref{e10}), we must choose
\be
\g_0 \eta_{(2\ell)}=i\eta_{(2\ell)}, ~~~ \bar \eta^{(2\ell)} \g_0=i\bar \eta^{(2\ell)}.
\ee
Then we get
\bea \label{j13}
&& \d \bar f_1^{(2\ell)}=-i\sr{\f{8\pi}{k}}\bar\eta^{(2\ell)}\th^{1\hi} \mD_0\phi_\hi^{(2\ell)}, ~~~
   \bar g_1^{(2\ell)}=-i\sr{\f{8\pi}{k}}\bar\eta^{(2\ell)}\th^{1\hi} \phi_\hi^{(2\ell)},      \nn\\
&& \d f_2^{(2\ell)}=i\sr{\f{8\pi}{k}}\bar\th_{1\hi}\eta_{(2\ell)} \mD_0\bar\phi^\hi_{(2\ell)}, ~~~
   g_2^{(2\ell)}=i\sr{\f{8\pi}{k}}\bar\th_{1\hi}\eta_{(2\ell)} \bar\phi^\hi_{(2\ell)} .
\eea
One can show that, given\footnote{We stress again that there are no summations of $\ell$ in this paper unless indicated explicitly.}
\be \label{j16}
\eta_{(2\ell)} \bar \eta^{(2\ell)}=-i-\g_{0},
\ee
equations (\ref{e10}) are satisfied. It is similar for conformal SUSY transformation. Thus we get the 1/2 BPS Wilson loop along a timelike infinite straight line.

\subsection{Relation between 1/4 and 1/2 BPS Wilson loops}

We check
\be
W_{1/2}-W_{1/4}=Q V,
\ee
for some supercharge $Q$ preserved by both $W_{1/4}$ and $W_{1/2}$ and some operator $V$.
This is similar to the ABJM case in \cite{Drukker:2009hy}.
In the $\mN=4$ SCSM theory we have the 1/4 and 1/2 BPS Wilson loops (\ref{w14ne4}) and (\ref{w12ne4})
\bea
&& W_{1/4}(s,t)=\mP \exp \lt( -i\int^s_t d\t L_{1/4}(\t) \rt), \nn\\
&& W_{1/2}(s,t)=\mP \exp \lt( -i\int^s_t d\t L_{1/2}(\t) \rt).
\eea
In this subsection it is convenient to rearrange the rows and columns and rewrite
\bea
&& L_{1/4}=\diag \lt( L_{1/4}^{(0)},L_{1/4}^{(1)},\cdots,L_{1/4}^{(n-1)} \rt), \nn\\
&& L_{1/4}^{(\ell)}=\lt( \ba{cc} \mA^{(2\ell+1)} & \\ & \hat\mA^{(2\ell)} \ea \rt), \nn\\
&& L_{1/2}=\diag \lt( L_{1/2}^{(0)},L_{1/2}^{(1)},\cdots,L_{1/2}^{(n-1)} \rt),\\
&& L_{1/2}^{(\ell)}=\lt( \ba{cc} \mA^{(2\ell+1)} & \bar f_1^{(2\ell)} \\ f_2^{(2\ell)} & \hat\mA^{(2\ell)} \ea \rt). \nn
\eea
Note that for $L_{1/4}$ there are $M^i_{\ph{i}j}=M^\hi_{\ph{\hi}\hj}=N_i^{\ph{i}j}=N_\hi^{\ph{\hi}\hj}=\diag (-1,1)$,
and for $L_{1/2}$ there are
$M^i_{\ph{i}j}=N_i^{\ph{i}j}=\diag (-1,1)$ and
$M^\hi_{\ph{\hi}\hj}=N_\hi^{\ph{\hi}\hj}=\diag (1,1)$.
Here $L_{1/2}^{(\ell)}$ with $\ell=0,1,\cdots,n-1$ can be thought as building blocks of the 1/2 BPS Wilson loop.
Note that though $L_{1/2}^{(\ell)}$ with fixed $\ell$ only involves
fermions  $\psi^i_{(2\ell)}$ and $\bar\psi_i^{(2\ell)}$ which are in (anti-)bifundamental representation of $U(N)_{(2\ell)}\times U(N)_{(2\ell+1)}$,
the involved scalar fields include not only $\phi_\hi^{(2\ell)},\bar\phi^\hi_{(2\ell)}$, but also
$\phi_i^{(2\ell\pm 1)}, \bar\phi^i_{(2\ell\pm 1)}$. In some sense, this construction is a kind of hybrid of 1/2 BPS Wilson loop in ABJM theory
and 1/4 BPS Wilson loop in  this $\mN=4$ theory.

We also define
\bea
&& L_{1/2}-L_{1/4}=\td  L=\td L_B+\td L_F, \nn\\
&& \td L_{1/2}=L_{1/4}+\td L_B-\td L_F,  \nn\\
&& \td W_{1/2}(s,t)=\mP \exp \lt( -i\int^s_t d\t \td L_{1/2}(\t) \rt).
\eea
Explicitly there are
\bea
&& \td L_B=\diag \lt( \td L_B^{(0)},\td L_B^{(1)},\cdots,\td L_B^{(n-1)} \rt),                           \nn\\
&& \td L_B^{(\ell)}=\f{4\pi}{k} \lt(\ba{cc} \phi_{\hat 1}^{(2\ell)}\bar\phi^{\hat 1}_{(2\ell)} &         \\
                                          & \bar\phi^{\hat 1}_{(2\ell)}\phi_{\hat 1}^{(2\ell)} \ea \rt), \nn\\
&& \td L_F=\diag \lt( \td L_F^{(0)},\td L_F^{(1)},\cdots,\td L_F^{(n-1)} \rt),          \\
&& \td L_F^{(\ell)}=\sr{\f{2\pi}{k}} \lt(\ba{cc}  & \bar\eta^{(2\ell)}\psi^1_{(2\ell)}  \\
                                   \bar\psi_1^{(2\ell)}\eta_{(2\ell)} & \ea \rt).       \nn
\eea
Then we get
\bea
&& W_{1/2}(s,t)-W_{1/4}(s,t)=-i \int^s_t d\t \lt[ W_{1/4}(s,\t) \td L(\t) W_{1/2}(\t,t)  \rt]  \nn\\
&& \phantom{W_{1/2}(s,t)-W_{1/4}(s,t)}
                           =-i \int^s_t d\t \lt[ W_{1/2}(s,\t) \td L(\t) W_{1/4}(\t,t) \rt].
\eea
We define
\bea
&& \L=\diag \lt( \L^{(0)},\L^{(1)},\cdots,\L^{(n-1)} \rt) ,\nn\\
&& \L^{(\ell)}=-\sr{\f{2\pi}{k}} \lt(\ba{cc} & \phi_{\hat 1}^{(2\ell)} \\  \bar\phi^{\hat 1}_{(2\ell)} & \ea \rt),
\eea
and a Grassmann odd operator
\bea
&& Q=\diag \lt( Q^{(0)},Q^{(1)},\cdots,Q^{(n-1)} \rt) ,\nn\\
&& Q^{(\ell)}=\bar\eta^{(2\ell)} P^{1\hat 1}+\bar P_{1\hat 1}\eta_{(2\ell)},
\eea
with $P^{1\hat 1}$ and $\bar P_{1\hat 1}$ being Poncar\'e charges in (\ref{j14}).
It can be checked that
\be
Q \L=\td L_F, ~~~ \a\L^2=\td L_B, ~~~ \a=2.
\ee
Now we have the SUSY transformation
\bea
&& \d W_{1/4}(s,t)=0, \nn\\
&& \d W_{1/2}(s,t)=-i G(s)W_{1/2}(s,t)+W_{1/2}(s,t) i G(t), \nn\\
&& \d \td W_{1/2}(s,t)=i G(s) \td W_{1/2}(s,t)- \td W_{1/2}(s,t) i G(t),
\eea
with
\bea
&& G=\diag \lt( G^{(0)},G^{(1)},\cdots,G^{(n-1)} \rt) ,\nn\\
&& G^{(\ell)}=\lt( \ba{cc} & \bar g_1^{(2\ell)} \\ g_2^{(2\ell)} & \ea \rt).
\eea
Note that $\bar g_1^{(2\ell)}$ and $g_2^{(2\ell)}$ have been derived in (\ref{j13}).
And then we have
\bea
&& Q W_{1/4}(s,t)=0, \nn\\
&& Q W_{1/2}(s,t)=\a\L(s) W_{1/2}(s,t)- \td W_{1/2}(s,t) \a\L(t), \nn\\
&& Q \td W_{1/2}(s,t)=-\a\L(s) \td W_{1/2}(s,t)+  W_{1/2}(s,t) \a\L(t).
\eea
Note that from (\ref{j16}) we have $\bar \eta^{(2\ell)} \eta_{(2\ell)}=-2i$ with no summation of $\ell$.

We define
\bea
&& S_1(s,t)=-i \int^s_t d\t \lt[ W_{1/4}(s,\t) \L(\t) W_{1/2}(\t,t)  \rt], \nn\\
&& S_2(s,t)=-i \int^s_t d\t \lt[ W_{1/4}(s,\t) \L(\t) \td W_{1/2}(\t,t)  \rt], \nn\\
&& S_3(s,t)=-i \int^s_t d\t \lt[ W_{1/2}(s,\t) \L(\t) W_{1/4}(\t,t)  \rt],  \\
&& S_4(s,t)=-i \int^s_t d\t \lt[ \td W_{1/2}(s,\t) \L(\t) W_{1/4}(\t,t)  \rt]. \nn
\eea
We can show that
\bea
&& Q S_1(s,t)=W_{1/2}(s,t)-W_{1/4}(s,t)-S_2(s,t)\a\L(t),  \nn\\
&& Q S_4(s,t)=W_{1/2}(s,t)-W_{1/4}(s,t)-\a\L(s)S_4(s,t).
\eea
For the infinite straight line, we have $s\to\inf$ and $t\to-\inf$, and we also assume $\L(\pm\inf)=0$.
Then we get
\be
W_{1/2}-W_{1/4}=Q S_1=Q S_4.
\ee
Operator $S_1$ or $S_4$ is just the $V$ we are looking for.

\section{Straight line in Euclidean space}\label{s4}

There are BPS Wilson loops along spacelike infinite straight lines in Euclidean space. Since they are similar to BPS Wilson loops along timelike infinite straight lines in Minkowski spacetime, and so it will be brief in this section.

\subsection{1/4 BPS Wilson loop}

We use coordinates $x^\m=(x^1,x^2,x^3)$ in Euclidean space. We have the 1/4 BPS Wilson loop along the infinite straight line $x^\m=\t\d^\m_1$ the same as (\ref{w14ne4}) except that
\be
M^i_{\ph{i}j}=M^\hi_{\ph{\hi}\hj}=N_i^{\ph{i}j}=N_\hi^{\ph{\hi}\hj}=\diag( i,-i).
\ee
The preserved Poncar\'e and conformal supersymmetries are
\bea
&& \g_1\th^{1\hat 1}=\th^{1\hat 1}, ~~~ \g_1\th^{2 \hat 2}=-\th^{2 \hat 2},  \nn\\
&& \g_1\vth^{1\hat 1}=\vth^{1\hat 1}, ~~~ \g_1\vth^{2 \hat 2}=-\vth^{2 \hat 2}, \\
&& \th^{1\hat 2}=\th^{2\hat 1}=\vth^{1\hat 2}=\vth^{2\hat 1}=0. \nn
\eea

\subsection{1/2 BPS Wilson loop}

Also we have the 1/2 BPS Wilson loop along the infinite straight line $x^\m=\t\d^\m_1$ the same as (\ref{w12ne4}) except that
\bea
&& M^i_{\ph{i}j}=N_i^{\ph{i}j}=\diag( i,-i), ~~~
   M^\hi_{\ph{\hi}\hj}=N_\hi^{\ph{\hi}\hj}=\diag( -i,-i),    \nn\\
&& \g_1 \eta_{(2\ell)}=\eta_{(2\ell)}, ~~~
   \bar \eta^{(2\ell)} \g_1=\bar \eta^{(2\ell)},~~~
   \eta_{(2\ell)} \bar \eta^{(2\ell)}=i(1+\g_{1}).
\eea
The preserved supersymmetries are
\bea
&& \g_1\th^{1\hi}=\th^{1\hi},      ~~~
\g_1\th^{2\hi}=-\th^{2\hi},         \nn\\
&& \g_1\vth^{1\hi}=\vth^{1\hi},      ~~~
\g_1\vth^{2\hi}=-\vth^{2\hi},
\eea
with $\hi=\hat 1,\hat2$.

\subsection{Relation between 1/4 and 1/2 BPS Wilson loops}

The check of $W_{1/2}-W_{1/4}=QV$ for a straight line in Euclidean space is similar to the case of a timelike straight line in Minkowski spacetime. The only differences are that
\bea
&& \td L_B^{(\ell)}=-\f{4\pi i}{k} \lt(\ba{cc} \phi_{\hat 1}^{(2\ell)}\bar\phi^{\hat 1}_{(2\ell)} & \\
                                          & \bar\phi^{\hat 1}_{(2\ell)}\phi_{\hat 1}^{(2\ell)} \ea \rt), \nn\\
&& \a=-2i.
\eea

\section{Circle in Euclidean space}\label{s5}

The $\mN=4$ SCSM theory is a superconformal theory, and a conformal transformation can change an infinite straight line to a circle. So there would be BPS circular Wilson loops if there exist BPS Wilson loops along infinite straight lines.

\subsection{1/4 BPS Wilson loop}

There is 1/4 circular BPS Wilson loop along $x^\m=(\cos\t,\sin\t,0)$ the same as (\ref{w14ne4}) except that
\be
M^i_{\ph{i}j}=M^\hi_{\ph{\hi}\hj}=N_i^{\ph{i}j}=N_\hi^{\ph{\hi}\hj}=\diag( i,-i).
\ee
The preserved Poncar\'e and conformal supersymmetries are
\bea
&& \vth^{1\hat1}=i\g_3\th^{1\hat1}, ~~~ \vth^{2\hat 2}=-i\g_3\th^{2\hat 2},   \nn\\
&& \th^{1\hat 2}=\th^{2\hat 1}=\vth^{1\hat 2}=\vth^{2\hat 1}=0.
\eea

\subsection{1/2 BPS Wilson loop}

Also there is circular 1/2 BPS Wilson loop along $x^\m=(\cos\t,\sin\t,0)$ the same as (\ref{w12ne4}) except that
\bea
&& M^i_{\ph{i}j}=N_i^{\ph{i}j}=\diag( i,-i), ~~~
   M^\hi_{\ph{\hi}\hj}=N_\hi^{\ph{\hi}\hj}=\diag( -i,-i), \nn \\
&& \bar\eta^{(2\ell)\a}=\bar\b(e^{i\t/2},e^{-i\t/2}),~~~
   \eta_{(2\ell)\a}=(e^{-i\t/2},e^{i\t/2})\b,
\eea
with $\b$, $\bar\b$ being Grassmann even constants and satisfying $\b\bar\b=i$. Note that we have  useful relations with no summations of $\ell$
\be
\eta_{(2\ell)} \bar\eta^{(2\ell)}=i \lt( 1+\dot x^\m \g_\m \rt), ~~~
\bar\eta^{(2\ell)} \eta_{(2\ell)}=2i.
\ee
Here we have antiperiodic boundary conditions
\be
G(2\pi)=-G(0),
\ee
and so the gauge invariant Wilson loop is
\be
\Tr W_{1/2}.
\ee
Now the preserved supersymmetries are
\be
\vth^{1\hi}=i\g_3\th^{1\hi}, ~~~ \vth^{2\hi}=-i\g_3\th^{2\hi},
\ee
with $\hi=\hat1,\hat2$.
$\phantom{\ref{}}$

\subsection{Relation between 1/4 and 1/2 BPS Wilson loops}\label{QVmain}

The check of $W_{1/2}-W_{1/4}=QV$ for a circle in Euclidean space is different to the case of a straight line in Minkowski spacetime. Firstly we have the differences
\bea\label{eqrelation}
&& \td L_B^{(\ell)}=-\f{4\pi i}{k} \lt(\ba{cc} \phi_{\hat 1}^{(2\ell)}\bar\phi^{\hat 1}_{(2\ell)} & \\
                                          & \bar\phi^{\hat 1}_{(2\ell)}\phi_{\hat 1}^{(2\ell)} \ea \rt), \nn\\
&& \L^{(\ell)}=-\sr{\f{2\pi}{k}} e^{i\t/2} \lt(\ba{cc} & \phi_{\hat 1}^{(2\ell)} \\  \bar\phi^{\hat 1}_{(2\ell)} & \ea \rt),  \\
&& Q^{(\ell)}=\bar\z^{(2\ell)} \lt( P^{1\hat 1} + i\g_3S^{1\hat 1} \rt)
             +\lt( \bar P_{1\hat 1} +\bar S_{1\hat 1} i\g_3 \rt) \z_{(2\ell)},  \nn\\\
&& \bar\z^{(2\ell)\a}=\bar\b(1,0), ~~~ \z_{(2\ell)\a}=(0,1)\b, ~~~  \a=-2ie^{-i\t}. \nn
\eea
Then, the construction of $V$ is also different, since we have to treat the boundary terms carefully. The calculation is very involved, and so we collect them in Appendix~\ref{QVappe}.

\section{Conclusion and discussion}\label{s6}

In this paper, we have investigated the supersymmetric Wilson loops in $\mN=4$ SCSM theory. In Minkowski spacetime we have 1/2 BPS Wilson loops along null infinite straight lines, and 1/4 and 1/2 BPS Wilson loops along timelike infinite straight lines. In Euclidean space we have 1/4 and 1/2 Wilson loops along infinite straight lines, as well as circular 1/4 and 1/2 Wilson loops. We also gave a complete proof that the difference between 1/4 and 1/2 Wilson loops is $Q$-exact with $Q$ being some supercharge that is preserved by both the 1/4 and 1/2 Wilson loops.
On the gravity side, we also studied the probe M2-branes dual to half BPS circular Wilson loops in the fundamental representation, and give the holographic prediction of the VEV of this 1/2 BPS Wilson loops in the M-theory limit.

The VEV of the half-BPS circular Wilson loop in fundamental representation can be calculated using localization in the M-theory limit ($N\to\infty$ with $k$ and $n$ being fixed) based on results in \cite{Herzog:2010hf}.
It will be also interesting to compute the vacuum expectation values of these BPS Wilson loops beyond the M-theory limit.
We can use the fermi gas approach \cite{Marino:2011eh, Klemm:2012ii} to include all of the $1/N$ corrections.
Similar to the ABJM case \cite{Hatsuda:2013yua}, it is interesting to study these Wilson loops in arbitrary representations.
These results are in accordance with the gravity ones, and they will be presented in \cite{Ouyang:2015hta}.

We think  that the construction of DT type BPS Wilson loops here could  be easily generalized to similar ones in  $\mN=4$ theories obtained from orbifolding ABJ theory or $\mN=5$ theories in \cite{Aharony:2008gk, Hosomichi:2008jb}.
It is an interesting question to study whether there exist DT type BPS Wilson loops
in other $\mN=4$ theories
\cite{Gaiotto:2008sd,Hosomichi:2008jd,Imamura:2008dt,Chen:2012at,Chen:2012bt}\footnote{This issue has been addressed recently in \cite{Cooke:2015ila}.}
and $\mN=3$ theories \cite{Jafferis:2008qz,Hohenegger:2009as,Gaiotto:2009tk,Hikida:2009tp}. As mentioned in the Introduction, in the latter case the DT type  Wilson loops are believed to be at most 1/3 BPS \cite{Chen:2014gta}.

\section*{Acknowledgments}

We would like to thank Bin Chen, Fa-Min Chen, Jiang Long, Jian-Xin Lu, Zohar Komargodski, Zhi-Guang Xiao and Meng-Qi Zhu for valuable discussions.
JW would like to thank KIAS and ICTS-USTC for hospitality during recent visits.
The work was in part supported by NSFC Grants  No.~11222549 and No.~11575202.
JW also gratefully acknowledges the support of K.~C.~Wong Education Foundation and Youth Innovation Promotion Association of CAS  (2011016).

\begin{appendix}

\section{Review of Wilson loops in ABJM theory}\label{abjmrev}

The ABJM theory is an $\mN=6$ SCSM theory, and it was constructed in \cite{Aharony:2008ug}.
ABJM theory has gauge group $U(N)\times U(N)$ and Chern-Simons levels $(k,-k)$, and the gauge fields are $A_\m$ and $\hat A_\m$ respectively.
The complex scalar $\phi_I$ and Dirac spinor $\psi_I$ are in $(N,\bar N)$ bifundamental representation,
and so $\bar\phi^I=\phi_I^\dagger$ and $\bar\psi_I=(\psi^I)^\dagger$ are in $(\bar N, N)$ representation.
We adopt the convention of spinors in three-dimensional Minkowski spacetime and Euclidean space in \cite{Ouyang:2015ada}.
We use $I,J,K,L,\cdots=1,2,3,4$ as indices of the $SU(4)$ R-symmetry.
A general SUSY transformation of ABJM theory is \cite{Gaiotto:2008cg,Hosomichi:2008jb,Terashima:2008sy,Bandres:2008ry}
\bea \label{abjmsusy}
&& \d A_\m=\f{4\pi}{k} \lt( \phi_I\bar\psi_J\g_\m\chi^{IJ} +\bar\chi_{IJ}\g_\m\psi^J\bar\phi^I \rt), \nn\\
&& \d\hat A_\m=\f{4\pi}{k} \lt( \bar\psi_J\g_\m\phi_I\chi^{IJ}+\bar\chi_{IJ}\bar\phi^I\g_\m\psi^J \rt), \nn\\
&& \d\phi_I=2i\bar\chi_{IJ}\psi^J, ~~~ \d\bar\phi^I=2i\bar\psi_J\chi^{IJ},\\
&& \d\psi^I=2\g^\m\chi^{IJ}D_\m\phi_J +2\vth^{IJ}\phi_J
            -\f{4\pi}{k}\chi^{IJ} \lt( \phi_J\bar\phi^K\phi_K-\phi_K\bar\phi^K\phi_J \rt)
            -\f{8\pi}{k}\chi^{KL}\phi_K\bar\phi^I\phi_L, \nn\\
&& \d\bar\psi_I=-2\bar\chi_{IJ}\g^\m D_\m\bar\phi^J +2\bar\vth_{IJ}\bar\phi^J
                +\f{4\pi}{k}\bar\chi_{IJ} \lt( \bar\phi^J\phi_K\bar\phi^K-\bar\phi^K\phi_K\bar\phi^J \rt)
                +\f{8\pi}{k}\bar\chi_{KL}\bar\phi^K\phi_I\bar\phi^L, \nn
\eea
with $\chi^{IJ}=\th^{IJ}+x^\m\g_\m\vth^{IJ}$ and $\bar\chi_{IJ}=\bar\th_{IJ}-\bar \vth_{IJ}x^\m\g_\m$.
The definitions of covariant derivatives are
\bea
&& D_\m\phi_J =\p_\m \phi_J +i A_\m \phi_J -i \phi_J \hat A_\m ,\nn\\
&& D_\m\bar\phi^J=\p_\m\bar\phi^J +i \hat A_\m \bar\phi^J -i \bar\phi^J  A_\m.
\eea
Also $\th^{IJ}$, $\bar\th_{IJ}$ and $\vth^{IJ}$, $\bar\vth_{IJ}$ are Dirac spinors with constraints
\bea \label{thij}
&& \th^{IJ}=-\th^{JI}, ~~~ (\th^{IJ})^*=\bar \th_{IJ}, ~~~ \bar\th_{IJ}=\f{1}{2}\e_{IJKL}\th^{KL}, \nn\\
&& \vth^{IJ}=-\vth^{JI}, ~~~ (\vth^{IJ})^*=\bar \vth_{IJ}, ~~~ \bar\vth_{IJ}=\f{1}{2}\e_{IJKL}\vth^{KL}.
\eea
Symbol $\e_{IJKL}$ is totally antisymmetric with $\e_{1234}=1$.
The $\th$, $\bar\th$ terms denote Poncar\'e SUSY transformation, and $\vth$, $\bar\vth$ terms denote conformal SUSY transformation. Note that we have $\d A_\m=\d A_\m^\dagger$, $\d \hat A_\m=\d \hat A_\m^\dagger$, $\d\bar\phi^I=\d\phi_I^\dagger$, and $\d\bar\psi^I=\d\psi_I^\dagger$.

For the Euclidean ABJM theory, the SUSY transformation is formally identical to (\ref{abjmsusy}), with $\chi^{IJ}=\th^{IJ}+x^\m\g_\m\vth^{IJ}$ and $\bar\chi_{IJ}=\bar\th_{IJ}-\bar \vth_{IJ}x^\m\g_\m$.
But now equations (\ref{thij}) become
\bea \label{z4}
&& \th^{IJ}=-\th^{JI}, ~~~ \bar\th_{IJ}=\f{1}{2}\e_{IJKL}\th^{KL}, \nn\\
&& \vth^{IJ}=-\vth^{JI}, ~~~ \bar\vth_{IJ}=\f{1}{2}\e_{IJKL}\vth^{KL}.
\eea
Note the twelve spinors $\th^{IJ}$, $\vth^{IJ}$ with $I,J=1,2,3,4$ are independent Dirac spinors.

In Minkowski spacetime, one has the 1/6 BPS Wilson loop along the timelike infinite straight line $x^\m=\t\d^\m_0$ \cite{Drukker:2008zx,Chen:2008bp,Rey:2008bh}
\bea \label{a5}
&& W_{1/6}=\mP \exp \lt( -i\int d\t \mA(\t) \rt),                \nn\\
&& \hat W_{1/6}=\mP \exp \lt( -i\int d\t \hat\mA (\t) \rt),         \nn\\
&& \mA=A_\m \dot x^\m +\f{2\pi}{k} M^I_{\ph{I}J} \phi_I\bar\phi^J |\dot x|,           \\
&& \hat\mA=\hat A_\m \dot x^\m +\f{2\pi}{k} N_I^{\ph{I}J} \bar\phi^I\phi_J |\dot x|,  \nn\\
&& M^I_{\ph{I}J}=N_I^{\ph{I}J}=\diag( -1,-1,1,1 ).                \nn
\eea
Here $W_{1/6}$ and $\hat W_{1/6}$ can be combined to give the 1/6 BPS Wilson loop
\bea \label{w16abjm}
&& W_{1/6}=\mP \exp \lt( -i\int d\t L_{1/6}(\t) \rt),             \nn\\
&& L_{1/6}=\lt( \ba{cc} \mA & \\ & \hat\mA \ea \rt).
\eea
The preserved Poncar\'e and conformal supersymmetries are
\bea
&& \g_0\th^{12}=i\th^{12}, ~~~ \g_0\th^{34}=-i\th^{34},  \nn\\
&& \th^{13}=\th^{14}=\th^{23}=\th^{24}=0,\\
&& \g_0\vth^{12}=i\vth^{12}, ~~~
   \g_0\vth^{34}=-i\vth^{34}, \nn\\
&& \vth^{13}=\th^{14}=\vth^{23}=\vth^{24}=0. \nn
\eea
Also one has the 1/2 BPS Wilson loop along the timelike infinite straight line $x^\m=\t\d^\m_0$ \cite{Drukker:2009hy}
\bea \label{w12abjm}
&& W_{1/2}=\mP \exp \lt( -i\int d\t L_{1/2}(\t) \rt),               \nn\\
&& L_{1/2}=\lt( \ba{cc} \mA &\bar f_1 \\ f_2 & \hat\mA \ea \rt),    \nn\\
&& \mA=A_\m \dot x^\m+\f{2\pi}{k} M^I_{\ph IJ} \phi_I\bar\phi^J |\dot x|,               \nn\\
&& \hat\mA=\hat A_\m \dot x^\m+\f{2\pi}{k} N_I^{\ph IJ}\bar\phi^I \phi_J |\dot x|,\\
&& M^I_{\ph{I}J}=N_I^{\ph{I}J}=\diag( -1,1,1,1 ),                  \nn\\
&& \bar f_1=\sr{\f{2\pi}{k}}\bar\eta_I\psi^I |\dot x|, ~~~
   f_2=\sr{\f{2\pi}{k}}\bar\psi_I\eta^I |\dot x|,                            \nn\\
&& \bar\eta_I=\bar\eta \d^1_I, ~~~
   \eta^I=\eta \d_1^I,                                              \nn\\
&& \g_0 \eta=i\eta, ~~~
   \bar \eta \g_0=i\bar \eta,~~~
   \eta \bar \eta=-i-\g_{0}.                                       \nn
\eea
The preserved Poncar\'e and conformal supersymmetries are
\bea
&& \g_0\th^{1i}=i\th^{1i},      ~~~
\g_0\th^{ij}=-i\th^{ij},        \nn\\
&& \g_0\vth^{1i}=i\vth^{1i},    ~~~
   \g_0\vth^{ij}=-i\vth^{ij},
\eea
with $i,j=2,3,4$.
One can use localization techniques to calculate the vacuum expectation value of the 1/6 BPS Wilson loop \cite{Kapustin:2009kz},
and in order to generalize this to the $1/2$ BPS Wilson loops \cite{Marino:2009jd,Drukker:2010nc} one needs the relation between 1/6 and 1/2 BPS Wilson loops \cite{Drukker:2009hy}
\be
W_{1/2}-W_{1/6}=Q V,
\ee
for some supercharge $Q$ preserved by both $W_{1/6}$ and $W_{1/2}$ and some operator $V$. Operator $V$ has been given in \cite{Drukker:2009hy}, and $W_{1/2}-W_{1/6}=QV$ has been checked for the first several orders.\footnote{In fact, the localization is used to compute the  BPS circular Wilson loops in the Euclidean space discussed in the following. The discussion on the relation of $W_{1/6}$ and $W_{1/2}$ here can be taken as a warm-up.}
Note that there is no spacelike BPS Wilson loop in Minkowski spacetime \cite{Ouyang:2015ada}.
One has 1/2 BPS Wilson loops along null infinite straight lines.

In Euclidean space, we use coordinates $x^\m=(x^1,x^2,x^3)$. One has the 1/6 BPS Wilson loop along the infinite straight line $x^\m= \t \d^\m_1$ the same as (\ref{w16abjm}) except that
\be
M^I_{\ph{I}J}=N_I^{\ph{I}J}=\diag( i,i,-i,-i).
\ee
The preserved Poncar\'e and conformal supersymmetries are
\bea
&& \g_1\th^{12}=\th^{12}, ~~~ \g_1\th^{34}=-\th^{34},  \nn\\
&& \th^{13}=\th^{14}=\th^{23}=\th^{24}=0,\\
&& \g_1\vth^{12}=\vth^{12}, ~~~
   \g_1\vth^{34}=-\vth^{34}, \nn\\
&& \vth^{13}=\th^{14}=\vth^{23}=\vth^{24}=0. \nn
\eea
Also one has the 1/2 BPS Wilson loop along the infinite straight line $x^\m= \t \d^\m_1$ the same as (\ref{w12abjm}) except that
\bea
&& M^I_{\ph{I}J}=N_I^{\ph{I}J}=\diag( i,-i,-i,-i ),                   \\
&& \g_1 \eta=\eta, ~~~
   \bar \eta \g_1=\bar \eta,~~~
   \eta \bar \eta=i(1+\g_{1}).                                        \nn
\eea
The preserved supersymmetries are
\bea
&& \g_1\th^{1i}=\th^{1i},      ~~~
\g_1\th^{ij}=-\th^{ij},        \nn\\
&& \g_1\vth^{1i}=\vth^{1i},    ~~~
   \g_1\vth^{ij}=-\vth^{ij},
\eea
with $i,j=2,3,4$.
The check of $W_{1/2}-W_{1/6}=QV$ for a straight line in Euclidean space is similar to the previous case.

Besides, in Euclidean space one has the circular 1/6 BPS Wilson loop along $x^\m=(\cos\t,\sin\t,0)$ the same as (\ref{w16abjm}) except that
\be
M^I_{\ph{I}J}=N_I^{\ph{I}J}=\diag( i,i,-i,-i).
\ee
The preserved Poncar\'e and conformal supersymmetries are
\bea
&& \vth^{12}=i\g_3\th^{12}, ~~~ \vth^{34}=-i\g_3\th^{34},   \nn\\
&& \th^{13}=\th^{14}=\th^{23}=\th^{24}=0,                   \\
&& \vth^{13}=\th^{14}=\vth^{23}=\vth^{24}=0.                \nn
\eea
Also one has the circular 1/2 BPS Wilson loop along $x^\m=(\cos\t,\sin\t,0)$ the same as (\ref{w12abjm}) except that
\bea
&& M^I_{\ph{I}J}=N_I^{\ph{I}J}=\diag( i,-i,-i,-i ),        \\
&& \bar\eta^\a=\bar\b(e^{i\t/2},e^{-i\t/2}),~~~
   \eta_\a=(e^{-i\t/2},e^{i\t/2})\b,                     \nn
\eea
with $\b$, $\bar\b$ being constants and satisfying $\b\bar\b=i$.
The preserved supersymmetries are
\be
\vth^{1i}=i\g_3\th^{1i}, ~~~ \vth^{ij}=-i\g_3\th^{ij},
\ee
with $i,j=2,3,4$.
There has been no general form of $V$ in the check of $W_{1/2}-W_{1/6}=QV$, but there are the first several orders in the expansion of $V$ in  \cite{Drukker:2009hy}.

\section{A simple proof of gauge covariance of Wilson lines}\label{susycov}

We have a general line in spacetime parameterized by $\t\in[t,s]$. For gauge field $A(\t)$ we define the Wilson line
\be
W(s,t)=\mP \exp \lt( -i\int^s_t d\t A(\t) \rt),
\ee
with $\mP$ being path-ordering. For a general infinitesimal gauge transformation
\be
\d A \equiv D_\t \L =\p_\t \L +i [A,\L],
\ee
the Wilson line transforms as
\be \label{dwst}
\d W(s,t)=-i\L(s)W(s,t)+W(s,t)i\L(t).
\ee
This gauge covariance of Wilson lines is well-known, one can see a complete proof in, for example, the textbook \cite{Peskin:1995ev}. Here we give a simple proof using induction.

For $n\geq0$ we define the symbols
\bea
&& X_n(s,t)= \int^s_t d\t_1 \int^{\t_1}_t d\t_2 \cdots \int^{\t_{n-1}}_t d\t_n A_1A_2\cdots A_n, \nn\\
&& Y_n(s,t)= \int^s_t d\t_1 \int^{\t_1}_t d\t_2 \cdots \int^{\t_{n-1}}_t d\t_n \lt( \p\L_1A_2\cdots A_n
                                                                                   +A_1\p\L_2A_3\cdots A_n \rt. \nn\\
&&\phantom{Y_n(s,t)= \int^s_t d\t_1 \int^{\t_1}_t d\t_2 \cdots \int^{\t_{n-1}}_t d\t_n \lt(\rt.} \lt.
                                                                                   +\cdots
                                                                                   +A_1A_2\cdots A_{n-1}\p\L_n \rt), \\
&& Z_n(s,t)= \int^s_t d\t_1 \int^{\t_1}_t d\t_2 \cdots \int^{\t_{n-1}}_t d\t_n \lt( [A_1,\L_1]A_2\cdots A_n
                                                                                   +A_1[A_2,\L_2]A_3\cdots A_n \rt. \nn\\
&& \phantom{Z_n(s,t)= \int^s_t d\t_1 \int^{\t_1}_t d\t_2 \cdots \int^{\t_{n-1}}_t d\t_n \lt[\rt.} \lt.
                                                                                   +\cdots
                                                                                   +A_1A_2\cdots A_{n-1}[A_n,\L_n] \rt), \nn
\eea
with the shorthand $A_i\equiv A(\t_i)$, $\L_i\equiv \L(\t_i)$ and $\p \L_i\equiv \p_{\t_i}\L(\t_i)$. Note that we have $X_0=1$ and $Y_0=Z_0$=0. We have the relations
\be
W=\sum_{n=0}^{+\inf} (-i)^n X_n, ~~~ \d X_n=Y_n+ i Z_n.
\ee
With the recursive relations for $n\geq1$
\bea
&& X_n(s,t)=\int_t^s d\t A(\t)X_{n-1}(\t,t), \nn\\
&& Y_n(s,t)=\int_t^s d\t \lt( \p_\t\L(\t)X_{n-1}(\t,t)+A(\t)Y_{n-1}(\t,t) \rt),  \\
&& Z_n(s,t)=\int_t^s d\t \lt( [A(\t),\L(\t)]X_{n-1}(\t,t)+A(\t)Z_{n-1}(\t,t) \rt), \nn
\eea
we can use induction to prove
\be
Y_{n+1}(s,t)-Z_n(s,t)=\L(s)X_n(s,t)-X_n(s,t)\L(t).
\ee
This leads to the infinitesimal version of the gauge covariance of the Wilson line (\ref{dwst}).

We rewrite (\ref{dwst}) as
\be
\mP \lt( e^{-i\int^s_t d\t A(\t) } \int_t^s d {\t'} D_{\t'}\L(\t')\rt)=
\mP \lt( e^{-i\int^s_t d\t A(\t) } [\L(s)-\L(t)]\rt).
\ee
Then it follows that more generally for $[t',s'] \subset [t,s]$ we can easily get
\be \label{j8}
\mP \lt( e^{-i\int^s_t d\t A(\t) } \cdots \int_{t'}^{s'} d {\t'} D_{\t'}\L(\t') \cdots \rt)=
\mP \lt( e^{-i\int^s_t d\t A(\t) }\cdots [\L(s')-\L(t')]\cdots \rt).
\ee
This is just the statement in \cite{Drukker:2009hy} that one can integrate out the covariant derivative $D=d+iA$ term in the presence of path ordered $\exp(-i\int A)$.

\section{Alternative Wilson loops for a super connection}\label{alternative}

In this appendix we explore alternative definitions of Wilson loops for a super connection. The result is that we find no nontrivial ones.

A super connection $L$ can be written as $L=B+F$ with Grassmann even part $B$ being block diagonal and Grassmann odd part $F$ being block off-diagonal
\be
B=\lt(\ba{cc} B_1 & \\ & B_2  \ea\rt), ~~~ F=\lt(\ba{cc} & F_1 \\ F_2 & \ea\rt).
\ee
When defining the path-ordering for the Grassmann odd part $F$ of the supermatrix $L$, we have ambiguities. We can do it as an ordinary matrix
\be
\mP F(\t_1)F(\t_2)= \lt\{ \ba{ll} F(\t_1) F(\t_2) & \t_1\geq\t_2 \\ F(\t_2) F(\t_1) & \t_1<\t_2 \ea \rt. ,
\ee
or we can define the super path-ordering as
\be \label{sp}
\mS\mP F(\t_1)F(\t_2)= \lt\{ \ba{ll} F(\t_1) F(\t_2) & \t_1\geq\t_2 \\ -F(\t_2) F(\t_1) & \t_1<\t_2 \ea \rt. .
\ee
Note that when acting on two Grassmann even matrices, or one even matrix and one odd matrix, $\mS\mP$ is no different with $\mP$. Only when acting on two odd matrices, $\mS\mP$ is different from $\mP$ as shown above.

We can define the Wilson loop along a super connection $L$ as that of an ordinary connection
\be \label{wstl}
W(s,t)=\mP \exp \lt( -i\int^s_t d\t L(\t) \rt),
\ee
and this is just what is done in main part of the paper. As shown in the last appendix, for a transformation
\be \label{dL}
\d L=\p \L+i[L,\L],
\ee
with $\L=\Sigma+\Xi$, $\Sigma$ being Grassmann even and $\Xi$ being odd, the Wilson loop transforms as
\be
\d W(s,t)=-i\L(s)W(s,t)+W(s,t)i\L(t).
\ee
This is the same as the case of an ordinary gauge field. For super matrices there is also definition
\be
[L,\L\}=[B,\S]+[B,\Xi]+[F,\Xi]+\{F,\Xi\}.
\ee
Note that for a transformation
\be \label{dpL}
\d' L=\p \L+i[L,\L\}=\d L+2i\Xi F,
\ee
the Wilson loop (\ref{wstl}) transforms as
\be
\d' W(s,t)=-i\L(s)W(s,t)+W(s,t)i\L(t)+ 2 \mP\lt[ \exp \lt( -i\int^s_t d\t L(\t) \rt) \int^s_t d\t' \Xi(\t')F(\t') \rt],
\ee
and so it is not covariant under $\d'L$.

Alternatively, we can define the Wilson loop for a super connection $L$ as
\be \label{swstl}
SW(s,t)=\mS\mP \exp \lt( -i\int^s_t d\t L(\t) \rt),
\ee
with the super path-ordering defined in (\ref{sp}). We rewrite $SW$ as
\be
SW(s,t)=\mP \lt[ \exp \lt( -i\int^s_t d\t B(\t) \rt) \mS\mP \exp \lt( -i\int^s_t d\t F(\t) \rt) \rt],
\ee
and then expand
\be
\mS\mP \exp \lt( -i\int^s_t d\t F(\t) \rt)=\sum_{n=0}^{+\inf} \f{(-i)^n}{n!} T_n(s,t),
\ee
with
\be
T_n(s,t)= \mS\mP \int^s_t d\t_1 \int^{\t_1}_t d\t_2 \cdots \int^{\t_{n-1}}_t d\t_n
                  \sum_{\s\in S_n} F(\t_{\s(1)}) F(\t_{\s(2)})\cdots F(\t_{\s(n)}),
\ee
where $S_n$ denotes order $n$ permutation group. It is easy to see that $T_n=0$ for $n\geq2$. And then we get
\be
SW(s,t)=\mP \lt[ \exp \lt( -i\int^s_t d\t B(\t) \rt) \lt( 1 -i\int^s_t d\t F(\t) \rt) \rt].
\ee
Thus the definition of $SW$ (\ref{swstl}) is trivial for several aspects.
\begin{itemize}
  \item For a loop, upon taken the trace $\Tr$ or super trace $\STr$, there is no contribution from the block off-diagonal part,
        \bea
        && \Tr SW= \Tr \mP \exp \lt( -i \oint d\t B(\t) \rt), \nn\\
        && \STr SW= \STr \mP \exp \lt( -i \oint d\t B(\t) \rt).
        \eea
  \item Quantum mechanically, the Grassmann odd part will not contribute to the vacuum expectation value,
        \be
        \lag SW \rag= \lt\lag \mP\exp \lt( -i \int d\t B(\t) \rt) \rt\rag.
        \ee
  \item Furthermore, it is not covariant under the transformation (\ref{dL}) or (\ref{dpL}).
\end{itemize}

\section{A complete proof of 1/2 and 1/4 BPS Wilson loops differencing by a $Q$-exact term}\label{QVappe}

The result here is general and applies not only to the $\mN=4$ SCSM case, but also to the ABJM case. A complete proof of 1/2 and 1/6 BPS Wilson loops difference in ABJM theory being $Q$-exact is the same as what is presented here.

\subsection{Some simplifications}

First of all, let us repeat the problem that we are going to tackle and make some simplifications.
We have a circle parameterized by $\t\in[0,1]$ with $x^\m(1)=x^\m(0)$.\footnote{Note that this is different to what we have done before. For a circle we have used $\t \in [0,2\pi]$.}
We have the 1/4 and 1/2 BPS Wilson loops
\bea
&& W_{1/4}=\mP \exp \lt( -i \int d\t L_{1/4}(\t) \rt),  \nn\\
&& W_{1/2}=\mP \exp \lt( -i \int d\t L_{1/2}(\t) \rt),
\eea
with
\be
L_{1/2}=L_{1/4}+\td L=L_{1/4} +\td L_B +\td L_F.
\ee
Here $L_{1/2}$ is a supermatrix, with $L_{1/4} +\td L_B$ being its Grassmann even block diagonal part and $\td L_F$ being its Grassmann odd block off-diagonal part. As shown in (\ref{eqrelation}), we can find a Grassmann odd operator $Q$ and a Grassmann even block off-diagonal matrix $\L$ satisfying
\bea
&& Q\L=\td L_F, ~~~ \k \L^2=\td L_B, ~~~\k(1)=\k(0) , ~~~ \L(1)=-\L(0),  \\
&& Q L_{1/4}=0, ~~~ Q \td L_B=\{ \td L_F,\k\L \}, ~~~
   Q \td L_F=\p_\t(i\k\L) +i[ L_{1/4}, i\k\L ].       \nn
\eea
Here the factor $\a$ has been redefined as $\k$. Note that $[\td L_B,\L]=0$ has been used. We want to find some operators $V$ and $U$ that satisfy
\be
Q V(1,0)=W_{1/2}(1,0)-W_{1/4}(1,0) +i\k(1)\L(1) U(1,0) +U(1,0)i\k(0)\L(0).
\ee
Taking the trace we would have
\be
Q \Tr V(1,0)= \Tr W_{1/2}(1,0)-\Tr W_{1/4}(1,0).
\ee
We call this task~I.

To avoid cluster of factors and indices, we make the following redefinitions
\bea
&& L_{1/4} \to i L_{1/4}, ~~~ L_{1/2} \to i L_{1/2}, ~~~ \td L \to i L, ~~~ \td L_B \to i B, \nn\\
&&  \td L_F \to i F, ~~~\L \to e^{i\pi/4} \L, ~~~ Q \to e^{i\pi/4} Q.
\eea
Then we have
\bea
&& W_{1/4}=\mP \exp \lt( \int d\t L_{1/4}(\t) \rt),  \nn\\
&& W_{1/2}=\mP \exp \lt( \int d\t L_{1/2}(\t) \rt),
\eea
with
\be
L_{1/2}=L_{1/4}+ L=L_{1/4} +B +F.
\ee
We have the relations
\bea
&& Q\L=F, ~~~ \k\L^2= B, ~~~ \k(1)=\k(0),  ~~~ \L(1)=-\L(0),  \\
&& Q L_{1/4}=0, ~~~ Q B= \{ F,\k\L \},   ~~~ Q F= \p_\t(\k\L)-[ L_{1/4},  \k\L ].   \nn
\eea
Our task is still to find some operators $V$ and $U$ that satisfy
\be
Q V(1,0)=W_{1/2}(1,0)-W_{1/4}(1,0) +\k(1)\L(1) U(1,0) +U(1,0)\k(0)\L(0).
\ee
We call this task~II. Of course, it is equivalent to task~I.

Furthermore, we may set $L_{1/4}=0$ and redefine $W_{1/2}=W$ in task~II. Now there are
\bea \label{hahaha}
&& W=\mP \exp \lt( \int d\t L(\t) \rt), \nn\\
&& L=B +F.
\eea
We have the relations
\bea \label{j1}
&& Q\L=F, ~~~ \k\L^2= B, ~~~ \k(1)=\k(0), ~~~ \L(1)=-\L(0),  \nn\\
&& Q B= \{ F,\k\L \}, ~~~ Q F= \p_\t(\k\L).
\eea
Our task is to find some operators $V$ and $U$ that satisfy
\be \label{taskIII}
Q V(1,0)=W(1,0)-1+\k(1)\L(1)U(1,0) +U(1,0)\k(0)\L(0).
\ee
We call it task~III. It is a special case of task~II, and so is easier.

\subsection{Some definitions}

Before tacking task~III we make some formal definitions. We have a circle parameterized by $\t\in[0,1]$ with $x^\m(1)=x^\m(0)$, and we will also use $s,t,\t_1,\t_2,\cdots$ to denote the parameter of the circle. We define two kinds of quantities on the circle.
\begin{itemize}
  \item We call the first kind type~1, and a type~1 quantity has only one argument. Generally we denote them by lowercase Latin letters $a(\t),b(\t),c(\t),\cdots$, or simply $a,b,c,\cdots$. Type~1 quantities below will include $\k$, $\L$, $B$, $F$, et~al.
  \item The second kind is type~2, and a type~2 quantity has two arguments. We denote them generally by lowercase Greek letters $\a(s,t),\b(s,t),\g(s,t),\cdots$, or simply $\a,\b,\c,\cdots$. Note that $s \geq t$ is required. Type~2 quantities below will include $W$, $W_n$, $V$, $V_n$, $U$, $U_n$, $\L_{mn}$, $B_{mn}$, $F_{mn}$, $S_{5,6,7,8}$, et~al. We also define the identity type~2 quantity $I(s,t)=1$.
\end{itemize}

We then define two kinds of operations $*$ and $\circ$. For two type~1 quantities $a,b$ we define a type~2 quantity as
\be
(a*b)(s,t) \equiv \int^s_t d\t_1  \int^{\t_1}_t d\t_2 a(\t_1) b(\t_2)
                 =\int^s_t d\t_2  \int^s_{\t_2} d\t_1 a(\t_1) b(\t_2).
\ee
Note that $a*b\neq b*a$.
For one type~1 quantity $a$ and one type~2 quantity $\a$ we define type~2 quantities as
\bea
&& (\a*a)(s,t) \equiv \int^s_t d\t \a(s,\t) a(\t),  \nn\\
&& (a*\a)(s,t) \equiv \int^s_t d\t a(\t) \a(\t,t).
\eea
Note that $\a*a\neq a*\a$.
For two type~2 quantities $\a,\b$ we do NOT define $\a*\b$, and so it is illegal.
For one type~1 quantity $a$ and two type~2 quantity $\a,\b$, we define a type~2 quantity
\be
(\a\circ a \circ \b)(s,t) \equiv \int^s_t d\t \a(s,\t) a(\t) \b(\t,t).
\ee
Note that symbol $\circ$ must appear in pair. We also define shorthand
\be \label{shorthands}
(ab)(\t) \equiv a(\t)b(\t), ~~~
(a\a)(s,t) \equiv a(s)\a(s,t), ~~~
(\a a)(s,t) \equiv \a(s,t)a(t).
\ee
Note that the shorthand is of the highest priority in calculation.

Under the above definitions there are some useful relations. There are
\bea
&& a*I=I*a, ~~~ (I*a)*b=a*b, ~~~ a*(I*b)=a*b, \nn\\
&& I\circ a \circ \a=a*\a, ~~~ \a\circ a \circ I=\a*a.
\eea
Note that $I*a \neq a$, because they are of different types. We can prove the associative relations
\be
(a*b)*c=a*(b*c), ~~~ (a*\a)*b=a*(\a*b).
\ee
Also we know $a*(b*\a)$ and $(\a*a)*b$ are legal, and $(a*b)*\a$ and $\a*(a*b)$ are illegal. So we can write without ambiguity
\be
a*b*c, ~~~ a*\a*b, ~~~ a*b*\a, ~~~ \a*a*b.
\ee
We can prove
\be
a*(\a\circ b \circ \b) = (a*\a) \circ b \circ \b, ~~~
(\a\circ a \circ \b)*b = \a \circ a \circ (\b *b),
\ee
and so we can write directly
\be
a*\a \circ b \circ \b, ~~~
   \a\circ a \circ \b*b.
\ee
For the shorthand (\ref{shorthands}) we have useful relations
\bea
&& \a*ab=\a a*b, ~~~ ab*\a=a*b\a, \nn\\
&& \a a\circ b \circ \b=\a \circ ab \circ \b
                       =\a \circ a \circ b\b.
\eea

\subsection{The main part}

We tackle task~III in this subsection and Subsections \ref{branch1} and \ref{branch2}.
As done in \cite{Drukker:2009hy}, we expand (\ref{hahaha}) in powers of $B$ and $F$. We set $F$ to be of order one, and $B$ to be of order two. And then we have
\be \label{j5}
W=\sum_{n=0}^{+\inf} W_n,
\ee
with the first few orders being
\bea
&& W_0=I, ~~~ W_1=I*F, ~~~ W_2=I*B+F*F, ~~~
  W_3=B*F +F*B +F*F*F,  \nn\\
&& W_4=B*B +B*F*F +F*B*F +F*F*B +F*F*F*F.
\eea
It can be seen that our definitions simplify these formulas significantly.
There is recursive relation for $n\geq2$
\be \label{wnrec}
W_n=B*W_{n-2}+F*W_{n-1}=W_{n-2}*B+W_{n-1}*F.
\ee
Note that we can define $W_{-1}=0$ to make the above equations apply to $n \geq 1$.

We also define
\be
V=\sum_{n=1}^{+\inf} V_n, ~~~ U=\sum_{n=1}^{+\inf} U_n.
\ee
And then task~III (\ref{taskIII}) becomes to find $V_n$ and $U_n$ for $n\geq1$ to satisfy
\be \label{j2}
Q V_n=W_n+\k\L U_n +U_n\k\L.
\ee
We define
\be \label{j9}
\td W=\mP \exp \lt[ \int d\t \lt( B -F \rt) \rt],
\ee
and then we get
\be  \label{j6}
\td W=\sum_{n=0}^{+\inf} (-)^n W_n.
\ee
From (\ref{j1}) we have
\be \label{j4}
Q W =\k\L  W  - \td W \k\L , ~~~
Q \td W =-\k\L \td W  +  W \k\L ,
\ee
and then for $n\geq 1$ we get
\be
Q W_n =\k\L W_{n-1}+(-)^n W_{n-1}\k\L.
\ee
Note that $W_0=I$, $Q W_0=0$, and we have set $W_{-1}=0$. And then the above equation applies to $n \geq 0$.

For $m\geq-1$ and $n \geq -1$ we define
\be \label{j11}
\L_{mn} \equiv W_m \circ \L \circ W_n, ~~~
B_{mn} \equiv W_m \circ B \circ W_n, ~~~
F_{mn} \equiv W_m \circ F \circ W_n.
\ee
Note that we have $\L_{mn}=0$ for $m=-1$ or $n=-1$, and it is similar to $B_{mn}$ and $F_{mn}$.
Because of (\ref{wnrec}), for $m\geq1$ we have
\bea \label{BFrec}
&& B_{mn}=B*B_{m-2,n}+F*B_{m-1,n},  \nn\\
&& F_{mn}=B*F_{m-2,n}+F*F_{m-1,n},
\eea
and for $n\geq1$ we have
\bea
&& B_{mn}=B_{m,n-2}*B+B_{m,n-1}*F,  \nn\\
&& F_{mn}=F_{m,n-2}*B+F_{m,n-1}*F.
\eea
And we can show
\be \label{j3}
Q \L_{mn}=(-)^m \lt( B_{m-1,n} +B_{m,n-1}+F_{mn}  \rt) +\k\L \L_{m-1,n} +(-)^{m+n} \L_{m,n-1} \k\L.
\ee
Then we claim that the $V_n$ and $U_n$ with $n\geq1$ we want in (\ref{j2}) are that for $k\geq0$
\bea
&& V_{2k+1}=\sum_{i=0}^{2k} \L_{i,2k-i}, ~~~
   U_{2k+1}=\sum_{i=0}^{2k-1} \L_{i,2k-1-i},  \nn\\
&& V_{2k+2}=\f{1}{2k+2} \sum_{i=0}^{2k+1} (-)^i \L_{i,2k+1-i},  \\
&& U_{2k+2}=-\f{1}{2k+2} \sum_{i=0}^{2k} (-)^i \L_{i,2k-i}.  \nn
\eea
We will prove this claim using two different methods in Subsections~\ref{branch1} and \ref{branch2}.
The first few orders are
\bea
&& V_1=I*\L, ~~~ U_1=0, ~~~
   V_2=\f{1}{2}(\L*F-F*\L), ~~~ U_2=-\f{1}{2}I*\L,                          \nn\\
&& V_3=\L*B +B*\L +\L*F*F +F*\L*F +F*F*\L, ~~~
   U_3=\L*F+F*\L, \nn\\
&& V_4=\f{1}{4}             ( \L*B*F -F*B*\L +B*\L*F-B*F*\L+\L*F*B-F*\L*B   \\
&& \phantom{V_4=\f{1}{4} }    +\L*F*F*F -F*\L*F*F +F*F*\L*F -F*F*F*\L),     \nn\\
&& U_4=-\f{1}{4} (\L*B+B*\L   +\L*F*F -F*\L*F +F*F*\L ).                    \nn
\eea
Note that $V_{1,2,3}$ are just the ones given in \cite{Drukker:2009hy}. Here we have gone further, and give a general expression of $V_n$ and $U_n$ for all integers $n\geq 1$. Besides, we will give a general proof of (\ref{j2}).
There are two methods of doing so, and then we split the main part to two branches.

\subsection{The first branch}\label{branch1}

In the first branch we use (\ref{j3}) and get for $k\geq0$
\bea
&& Q V_{2k+1} = \sum_{i=0}^{2k} (-)^i F_{i,2k-i}  +\k\L U_{2k+1} + U_{2k+1}\k\L, \\
&& Q V_{2k+2} = \f{1}{2k+2} \lt( 2 \sum_{i=0}^{2k} B_{i,2k-i} +\sum_{i=0}^{2k+1}  F_{i,2k+1-i} \rt)
               +\k\L U_{2k+2} + U_{2k+2}\k\L. \nn
\eea
Thus to prove (\ref{j2}) we need for $k\geq0$
\bea \label{j7}
&& \sum_{i=0}^{2k} (-)^i F_{i,2k-i}=W_{2k+1},  \\
&& 2 \sum_{i=0}^{2k} B_{i,2k-i} +\sum_{i=0}^{2k+1}  F_{i,2k+1-i} =(2k+2)W_{2k+2}, \nn
\eea
the first few orders of which can be verified easily. Furthermore, for $k\geq0$ we can prove the above two equations and the following two equations
\bea
&& \sum_{i=0}^{2k+1} (-)^i F_{i,2k+1-i}=0,  \\
&& 2 \sum_{i=0}^{2k+1} B_{i,2k+1-i} +\sum_{i=0}^{2k+2}  F_{i,2k+2-i} =(2k+3)W_{2k+3}, \nn
\eea
using induction. In the process (\ref{wnrec}) and (\ref{BFrec}) are used. Thus the proof of (\ref{j2}) is done.

\subsection{The second branch}\label{branch2}

In the second branch, we firstly define
\bea \label{j10}
&& S_5= W \circ \L \circ W, ~~~ S_6= \td W \circ \L \circ \td W, \nn\\
&& S_7= \td W \circ \L \circ W, ~~~ S_8= W \circ \L \circ \td W.
\eea
Then we can show
\bea
&& S_5+S_6=2\sum_{k=0}^{+\inf}V_{2k+1}, ~~~ S_5-S_6=2\sum_{k=0}^{+\inf}U_{2k+1},  \nn\\
&& S_7+S_8=-2\sum_{k=0}^{+\inf} (2k+2)U_{2k+2}, ~~~ S_7-S_8=2\sum_{k=0}^{+\inf} (2k+2)V_{2k+2}.
\eea
Note that $U_1=0$ has been used.
From (\ref{j4}) we get
\bea
&& Q S_5= \td W \circ F \circ W+\k\L S_5-S_6 \k\L,  \nn\\
&& Q S_6=  W \circ F \circ \td W-\k\L S_6+S_5 \k\L,  \\
&& Q S_7=  W \circ (2B+F) \circ  W -\k\L S_7-S_8 \k\L,  \nn\\
&& Q S_8=  \td W \circ (-2B+F) \circ \td W+\k\L S_8+S_7 \k\L. \nn
\eea
And then using (\ref{j5}) and (\ref{j6}), as well the results (\ref{j7}) in the first branch, we can get
\bea
&& Q(S_5+S_6)=2 \sum_{k=0}^{+\inf} W_{2k+1}+\k\L (S_5-S_6)+ (S_5-S_6)\k\L,  \nn\\
&& Q(S_7-S_8)=2 \sum_{k=0}^{+\inf} (2k+2)W_{2k+2}   -\k\L (S_7+S_8) -(S_7+S_8)\k\L.
\eea
Thus for $k\geq0$ we get
\bea
&& Q V_{2k+1}=W_{2k+1}+\k\L U_{2k+1} +U_{2k+1}\k\L,  \nn\\
&& Q V_{2k+2}=W_{2k+2}+\k\L U_{2k+2} +U_{2k+2}\k\L.
\eea
This is indeed just (\ref{j2}). However, we note that neither $V$ nor $U$ can be written directly as combinations of $S_{5,6,7,8}$.

\subsection{Back to the main part}

Now we turn back to the main part, with task~III being completed. From task~III to task~II we need to use (\ref{j8}) and just make some simple changes in the proof of task~III. All the type~1 quantities do not change, and every type~2 quantity changes as
\be \label{j12}
\a(s,t)=\mP\lt[ W_{1/4}(s,t)\ul\a(s,t) \rt],
\ee
with $\ul\a$ being the old type~2 quantity and $\a$ being the new one. Especially, the ``identity'' type~2 quantity becomes
\be
I(s,t) \to  W_{1/4}(s,t).
\ee
For type~1 quantities $a,b,c,\cdots$ and the old type~2 quantities $\ul\a,\ul\b,\ul\g,\cdots$, there are still $*$ and $\circ$ operations defined as before. For type~1 quantities $a,b,c,\cdots$ and the new type~2 quantities $\a,\b,\g,\cdots$ we define $\cas$ and $\ccc$ operations as
\bea
&& (a\cas b)(s,t)  \equiv  \mP  \lt[ W_{1/4}(s,t) (a*b)(s,t)   \rt],    \nn\\
&& (a\cas \a)(s,t) \equiv \mP \lt[ W_{1/4}(s,t) (a*\ul\a)(s,t)  \rt],   \nn\\
&& (\a\cas a)(s,t) \equiv \mP \lt[ W_{1/4}(s,t)  (\ul\a*a)(s,t) \rt],   \\
&& (\a\ccc a \ccc \b)(s,t) \equiv \mP \lt[ W_{1/4}(s,t)  (\ul\a\circ a \circ \ul\b)(s,t) \rt]. \nn
\eea
Now we have shorthand
\bea
&& (a\a)(s,t) \equiv \mP  \lt[ W_{1/4}(s,t)a(s)\ul\a(s,t) \rt], \nn\\
&& (\a a)(s,t) \equiv \mP  \lt[ W_{1/4}(s,t)\ul\a(s,t)a(t) \rt].
\eea

Keeping the above changes in mind, we can tackle task~II with few efforts.
For example, we need to change (\ref{j5}) to
\be
W_{1/2}-W_{1/4}=\sum_{n=1}^{+\inf} W_n,
\ee
with $W_n$ being changed as (\ref{j12}).
And (\ref{j9}) is changed to
\be
\td W_{1/2}=\mP \exp \lt[ \int d\t \lt( L_{1/4}+ B -F \rt) \rt].
\ee
In the second branch equations (\ref{j10}) are changed to
\bea
&& S_5= W_{1/2} \ccc \L \ccc W_{1/2}, ~~~ S_6= \td W_{1/2} \ccc \L \ccc \td W_{1/2}, \nn\\
&& S_7= \td W_{1/2} \ccc \L \ccc W_{1/2}, ~~~ S_8= W_{1/2} \ccc \L \ccc \td W_{1/2}.
\eea
Thus task~II is completed following task~III and (\ref{j8}).
Since task~I is equivalent to task~II, task~I is competed too.

In summary we have given a complete proof that the difference of circular 1/2 and 1/4 BPS Wilson loops  in this $\mN=4$ SCSM theory is $Q$-exact, with $Q$ being some supercharge that is preserved by the both the 1/2 and 1/4 BPS Wilson loops.
As we have stated, this proof also applies to the 1/2 and 1/6 BPS Wilson loops in ABJM theory.

\end{appendix}


\begin{thebibliography}{10}

\bibitem{Maldacena:1997re}
J.~M. Maldacena, ``{The Large N limit of superconformal field theories and
  supergravity},'' {\em Adv.Theor.Math.Phys.} {\bfseries 2} (1998) 231--252,
\href{http://arxiv.org/abs/hep-th/9711200}{{\ttfamily arXiv:hep-th/9711200
  [hep-th]}}.

\bibitem{Gubser:1998bc}
S.~Gubser, I.~R. Klebanov, and A.~M. Polyakov, ``{Gauge theory correlators from
  noncritical string theory},''
  \href{http://dx.doi.org/10.1016/S0370-2693(98)00377-3}{{\em Phys.Lett.}
  {\bfseries B428} (1998) 105--114},
\href{http://arxiv.org/abs/hep-th/9802109}{{\ttfamily arXiv:hep-th/9802109
  [hep-th]}}.

\bibitem{Witten:1998qj}
E.~Witten, ``{Anti-de Sitter space and holography},'' {\em
  Adv.Theor.Math.Phys.} {\bfseries 2} (1998) 253--291,
\href{http://arxiv.org/abs/hep-th/9802150}{{\ttfamily arXiv:hep-th/9802150
  [hep-th]}}.

\bibitem{Aharony:2008ug}
O.~Aharony, O.~Bergman, D.~L. Jafferis, and J.~Maldacena, ``{N=6 superconformal
  Chern-Simons-matter theories, M2-branes and their gravity duals},''
  \href{http://dx.doi.org/10.1088/1126-6708/2008/10/091}{{\em JHEP} {\bfseries
  0810} (2008) 091},
\href{http://arxiv.org/abs/0806.1218}{{\ttfamily arXiv:0806.1218 [hep-th]}}.

\bibitem{Maldacena:1998im}
J.~M. Maldacena, ``{Wilson loops in large N field theories},''
  \href{http://dx.doi.org/10.1103/PhysRevLett.80.4859}{{\em Phys.Rev.Lett.}
  {\bfseries 80} (1998) 4859--4862},
\href{http://arxiv.org/abs/hep-th/9803002}{{\ttfamily arXiv:hep-th/9803002
  [hep-th]}}.

\bibitem{Rey:1998ik}
S.-J. Rey and J.-T. Yee, ``{Macroscopic strings as heavy quarks in large N
  gauge theory and anti-de Sitter supergravity},''
  \href{http://dx.doi.org/10.1007/s100520100799}{{\em Eur.Phys.J.} {\bfseries
  C22} (2001) 379--394},
\href{http://arxiv.org/abs/hep-th/9803001}{{\ttfamily arXiv:hep-th/9803001
  [hep-th]}}.

\bibitem{Berenstein:1998ij}
D.~E. Berenstein, R.~Corrado, W.~Fischler, and J.~M. Maldacena, ``{The Operator
  product expansion for Wilson loops and surfaces in the large N limit},''
  \href{http://dx.doi.org/10.1103/PhysRevD.59.105023}{{\em Phys.Rev.}
  {\bfseries D59} (1999) 105023},
\href{http://arxiv.org/abs/hep-th/9809188}{{\ttfamily arXiv:hep-th/9809188
  [hep-th]}}.

\bibitem{Drukker:1999zq}
N.~Drukker, D.~J. Gross, and H.~Ooguri, ``{Wilson loops and minimal
  surfaces},'' \href{http://dx.doi.org/10.1103/PhysRevD.60.125006}{{\em
  Phys.Rev.} {\bfseries D60} (1999) 125006},
\href{http://arxiv.org/abs/hep-th/9904191}{{\ttfamily arXiv:hep-th/9904191
  [hep-th]}}.

\bibitem{Pestun:2007rz}
V.~Pestun, ``{Localization of gauge theory on a four-sphere and supersymmetric
  Wilson loops},'' \href{http://dx.doi.org/10.1007/s00220-012-1485-0}{{\em
  Commun.Math.Phys.} {\bfseries 313} (2012) 71--129},
\href{http://arxiv.org/abs/0712.2824}{{\ttfamily arXiv:0712.2824 [hep-th]}}.

\bibitem{Drukker:2008zx}
N.~Drukker, J.~Plefka, and D.~Young, ``{Wilson loops in 3-dimensional N=6
  supersymmetric Chern-Simons Theory and their string theory duals},''
  \href{http://dx.doi.org/10.1088/1126-6708/2008/11/019}{{\em JHEP} {\bfseries
  0811} (2008) 019},
\href{http://arxiv.org/abs/0809.2787}{{\ttfamily arXiv:0809.2787 [hep-th]}}.

\bibitem{Chen:2008bp}
B.~Chen and J.-B. Wu, ``{Supersymmetric Wilson Loops in N=6 Super
  Chern-Simons-matter theory},''
  \href{http://dx.doi.org/10.1016/j.nuclphysb.2009.09.015}{{\em Nucl.Phys.}
  {\bfseries B825} (2010) 38--51},
\href{http://arxiv.org/abs/0809.2863}{{\ttfamily arXiv:0809.2863 [hep-th]}}.

\bibitem{Rey:2008bh}
S.-J. Rey, T.~Suyama, and S.~Yamaguchi, ``{Wilson Loops in Superconformal
  Chern-Simons Theory and Fundamental Strings in Anti-de Sitter Supergravity
  Dual},'' \href{http://dx.doi.org/10.1088/1126-6708/2009/03/127}{{\em JHEP}
  {\bfseries 0903} (2009) 127},
\href{http://arxiv.org/abs/0809.3786}{{\ttfamily arXiv:0809.3786 [hep-th]}}.

\bibitem{Drukker:2009hy}
N.~Drukker and D.~Trancanelli, ``{A Supermatrix model for N=6 super
  Chern-Simons-matter theory},''
  \href{http://dx.doi.org/10.1007/JHEP02(2010)058}{{\em JHEP} {\bfseries 1002}
  (2010) 058},
\href{http://arxiv.org/abs/0912.3006}{{\ttfamily arXiv:0912.3006 [hep-th]}}.

\bibitem{Griguolo:2012iq}
L.~Griguolo, D.~Marmiroli, G.~Martelloni, and D.~Seminara, ``{The generalized
  cusp in ABJ(M) N = 6 Super Chern-Simons theories},''
  \href{http://dx.doi.org/10.1007/JHEP05(2013)113}{{\em JHEP} {\bfseries 1305}
  (2013) 113},
\href{http://arxiv.org/abs/1208.5766}{{\ttfamily arXiv:1208.5766 [hep-th]}}.

\bibitem{Cardinali:2012ru}
V.~Cardinali, L.~Griguolo, G.~Martelloni, and D.~Seminara, ``{New
  supersymmetric Wilson loops in ABJ(M) theories},''
  \href{http://dx.doi.org/10.1016/j.physletb.2012.10.051}{{\em Phys.Lett.}
  {\bfseries B718} (2012) 615--619},
\href{http://arxiv.org/abs/1209.4032}{{\ttfamily arXiv:1209.4032 [hep-th]}}.

\bibitem{Kim:2013oza}
N.~Kim, ``{Supersymmetric Wilson loops with general contours in ABJM theory},''
  \href{http://dx.doi.org/10.1142/S0217732313501502}{{\em Mod.Phys.Lett.}
  {\bfseries A28} (2013) 1350150},
\href{http://arxiv.org/abs/1304.7660}{{\ttfamily arXiv:1304.7660 [hep-th]}}.

\bibitem{Bianchi:2014laa}
M.~S. Bianchi, L.~Griguolo, M.~Leoni, S.~Penati, and D.~Seminara, ``{BPS Wilson
  loops and Bremsstrahlung function in ABJ(M): a two loop analysis},''
  \href{http://dx.doi.org/10.1007/JHEP06(2014)123}{{\em JHEP} {\bfseries 1406}
  (2014) 123},
\href{http://arxiv.org/abs/1402.4128}{{\ttfamily arXiv:1402.4128 [hep-th]}}.

\bibitem{Correa:2014aga}
D.~H. Correa, J.~Aguilera-Damia, and G.~A. Silva, ``{Strings in $AdS_4 \times
  \mathbb{CP}^{3}$ Wilson loops in $\mathcal N=$6 super Chern-Simons-matter and
  bremsstrahlung functions},''
  \href{http://dx.doi.org/10.1007/JHEP06(2014)139}{{\em JHEP} {\bfseries 1406}
  (2014) 139},
\href{http://arxiv.org/abs/1405.1396}{{\ttfamily arXiv:1405.1396 [hep-th]}}.

\bibitem{Kapustin:2009kz}
A.~Kapustin, B.~Willett, and I.~Yaakov, ``{Exact Results for Wilson Loops in
  Superconformal Chern-Simons Theories with Matter},''
  \href{http://dx.doi.org/10.1007/JHEP03(2010)089}{{\em JHEP} {\bfseries 1003}
  (2010) 089},
\href{http://arxiv.org/abs/0909.4559}{{\ttfamily arXiv:0909.4559 [hep-th]}}.

\bibitem{Jafferis:2010un}
D.~L. Jafferis, ``{The Exact Superconformal R-Symmetry Extremizes Z},''
  \href{http://dx.doi.org/10.1007/JHEP05(2012)159}{{\em JHEP} {\bfseries 1205}
  (2012) 159},
\href{http://arxiv.org/abs/1012.3210}{{\ttfamily arXiv:1012.3210 [hep-th]}}.

\bibitem{Hama:2010av}
N.~Hama, K.~Hosomichi, and S.~Lee, ``{Notes on SUSY Gauge Theories on
  Three-Sphere},'' \href{http://dx.doi.org/10.1007/JHEP03(2011)127}{{\em JHEP}
  {\bfseries 1103} (2011) 127},
\href{http://arxiv.org/abs/1012.3512}{{\ttfamily arXiv:1012.3512 [hep-th]}}.

\bibitem{Marino:2009jd}
M.~Marino and P.~Putrov, ``{Exact Results in ABJM Theory from Topological
  Strings},'' \href{http://dx.doi.org/10.1007/JHEP06(2010)011}{{\em JHEP}
  {\bfseries 1006} (2010) 011},
\href{http://arxiv.org/abs/0912.3074}{{\ttfamily arXiv:0912.3074 [hep-th]}}.

\bibitem{Drukker:2010nc}
N.~Drukker, M.~Marino, and P.~Putrov, ``{From weak to strong coupling in ABJM
  theory},'' \href{http://dx.doi.org/10.1007/s00220-011-1253-6}{{\em
  Commun.Math.Phys.} {\bfseries 306} (2011) 511--563},
\href{http://arxiv.org/abs/1007.3837}{{\ttfamily arXiv:1007.3837 [hep-th]}}.

\bibitem{Bianchi:2013zda}
M.~S. Bianchi, G.~Giribet, M.~Leoni, and S.~Penati, ``{1/2 BPS Wilson loop in
  N=6 superconformal Chern-Simons theory at two loops},''
  \href{http://dx.doi.org/10.1103/PhysRevD.88.026009}{{\em Phys.Rev.}
  {\bfseries D88} (2013) 026009},
\href{http://arxiv.org/abs/1303.6939}{{\ttfamily arXiv:1303.6939 [hep-th]}}.

\bibitem{Bianchi:2013rma}
M.~S. Bianchi, G.~Giribet, M.~Leoni, and S.~Penati, ``{The 1/2 BPS Wilson loop
  in ABJ(M) at two loops: The details},''
  \href{http://dx.doi.org/10.1007/JHEP10(2013)085}{{\em JHEP} {\bfseries 1310}
  (2013) 085},
\href{http://arxiv.org/abs/1307.0786}{{\ttfamily arXiv:1307.0786 [hep-th]}}.

\bibitem{Griguolo:2013sma}
L.~Griguolo, G.~Martelloni, M.~Poggi, and D.~Seminara, ``{Perturbative
  evaluation of circular 1/2 BPS Wilson loops in N = 6 Super Chern-Simons
  theories},'' \href{http://dx.doi.org/10.1007/JHEP09(2013)157}{{\em JHEP}
  {\bfseries 1309} (2013) 157},
\href{http://arxiv.org/abs/1307.0787}{{\ttfamily arXiv:1307.0787 [hep-th]}}.

\bibitem{Gaiotto:2007qi}
D.~Gaiotto and X.~Yin, ``{Notes on superconformal Chern-Simons-Matter
  theories},'' \href{http://dx.doi.org/10.1088/1126-6708/2007/08/056}{{\em
  JHEP} {\bfseries 0708} (2007) 056},
\href{http://arxiv.org/abs/0704.3740}{{\ttfamily arXiv:0704.3740 [hep-th]}}.

\bibitem{Cooke:2015ila}
M.~Cooke, N.~Drukker, and D.~Trancanelli, ``{A profusion of $1/2$ BPS Wilson
  loops in $\mathcal{N}=4$ Chern-Simons-matter theories},''
  \href{http://dx.doi.org/10.1007/JHEP10(2015)140}{{\em JHEP} {\bfseries 1510}
  (2015) 140},
\href{http://arxiv.org/abs/1506.07614}{{\ttfamily arXiv:1506.07614 [hep-th]}}.

\bibitem{Aharony:2008gk}
O.~Aharony, O.~Bergman, and D.~L. Jafferis, ``{Fractional M2-branes},''
  \href{http://dx.doi.org/10.1088/1126-6708/2008/11/043}{{\em JHEP} {\bfseries
  0811} (2008) 043},
\href{http://arxiv.org/abs/0807.4924}{{\ttfamily arXiv:0807.4924 [hep-th]}}.

\bibitem{Hosomichi:2008jb}
K.~Hosomichi, K.-M. Lee, S.~Lee, S.~Lee, and J.~Park, ``{N=5,6 Superconformal
  Chern-Simons Theories and M2-branes on Orbifolds},''
  \href{http://dx.doi.org/10.1088/1126-6708/2008/09/002}{{\em JHEP} {\bfseries
  0809} (2008) 002},
\href{http://arxiv.org/abs/0806.4977}{{\ttfamily arXiv:0806.4977 [hep-th]}}.

\bibitem{Lee:2010hk}
K.-M. Lee and S.~Lee, ``{1/2-BPS Wilson Loops and Vortices in ABJM Model},''
  \href{http://dx.doi.org/10.1007/JHEP09(2010)004}{{\em JHEP} {\bfseries 1009}
  (2010) 004},
\href{http://arxiv.org/abs/1006.5589}{{\ttfamily arXiv:1006.5589 [hep-th]}}.

\bibitem{Chen:2014gta}
B.~Chen, J.-B. Wu, and M.-Q. Zhu, ``{Holographical Description of BPS Wilson
  Loops in Flavored ABJM Theory},''
  \href{http://dx.doi.org/10.1007/JHEP12(2014)143}{{\em JHEP} {\bfseries 1412}
  (2014) 143},
\href{http://arxiv.org/abs/1410.2311}{{\ttfamily arXiv:1410.2311 [hep-th]}}.

\bibitem{Gaiotto:2008sd}
D.~Gaiotto and E.~Witten, ``{Janus Configurations, Chern-Simons Couplings, And
  The theta-Angle in N=4 Super Yang-Mills Theory},''
  \href{http://dx.doi.org/10.1007/JHEP06(2010)097}{{\em JHEP} {\bfseries 1006}
  (2010) 097},
\href{http://arxiv.org/abs/0804.2907}{{\ttfamily arXiv:0804.2907 [hep-th]}}.

\bibitem{Hosomichi:2008jd}
K.~Hosomichi, K.-M. Lee, S.~Lee, S.~Lee, and J.~Park, ``{N=4 Superconformal
  Chern-Simons Theories with Hyper and Twisted Hyper Multiplets},''
  \href{http://dx.doi.org/10.1088/1126-6708/2008/07/091}{{\em JHEP} {\bfseries
  0807} (2008) 091},
\href{http://arxiv.org/abs/0805.3662}{{\ttfamily arXiv:0805.3662 [hep-th]}}.

\bibitem{Benna:2008zy}
M.~Benna, I.~Klebanov, T.~Klose, and M.~Smedback, ``{Superconformal
  Chern-Simons Theories and AdS(4)/CFT(3) Correspondence},''
  \href{http://dx.doi.org/10.1088/1126-6708/2008/09/072}{{\em JHEP} {\bfseries
  0809} (2008) 072},
\href{http://arxiv.org/abs/0806.1519}{{\ttfamily arXiv:0806.1519 [hep-th]}}.

\bibitem{Imamura:2008nn}
Y.~Imamura and K.~Kimura, ``{On the moduli space of elliptic
  Maxwell-Chern-Simons theories},''
  \href{http://dx.doi.org/10.1143/PTP.120.509}{{\em Prog.Theor.Phys.}
  {\bfseries 120} (2008) 509--523},
\href{http://arxiv.org/abs/0806.3727}{{\ttfamily arXiv:0806.3727 [hep-th]}}.

\bibitem{Terashima:2008ba}
S.~Terashima and F.~Yagi, ``{Orbifolding the Membrane Action},''
  \href{http://dx.doi.org/10.1088/1126-6708/2008/12/041}{{\em JHEP} {\bfseries
  0812} (2008) 041},
\href{http://arxiv.org/abs/0807.0368}{{\ttfamily arXiv:0807.0368 [hep-th]}}.

\bibitem{Ouyang:2015ada}
H.~Ouyang, J.-B. Wu, and J.-j. Zhang, ``{BPS Wilson Loops in Minkowski
  Spacetime and Euclidean Space},''
\href{http://arxiv.org/abs/1504.06929}{{\ttfamily arXiv:1504.06929 [hep-th]}}.

\bibitem{Herzog:2010hf}
C.~P. Herzog, I.~R. Klebanov, S.~S. Pufu, and T.~Tesileanu, ``{Multi-Matrix
  Models and Tri-Sasaki Einstein Spaces},''
  \href{http://dx.doi.org/10.1103/PhysRevD.83.046001}{{\em Phys.Rev.}
  {\bfseries D83} (2011) 046001},
\href{http://arxiv.org/abs/1011.5487}{{\ttfamily arXiv:1011.5487 [hep-th]}}.

\bibitem{Marino:2011eh}
M.~Marino and P.~Putrov, ``{ABJM theory as a Fermi gas},''
  \href{http://dx.doi.org/10.1088/1742-5468/2012/03/P03001}{{\em J.Stat.Mech.}
  {\bfseries 1203} (2012) P03001},
\href{http://arxiv.org/abs/1110.4066}{{\ttfamily arXiv:1110.4066 [hep-th]}}.

\bibitem{Klemm:2012ii}
A.~Klemm, M.~Marino, M.~Schiereck, and M.~Soroush,
  ``{Aharony-Bergman-Jafferis¨CMaldacena Wilson loops in the Fermi gas
  approach},'' \href{http://dx.doi.org/10.5560/ZNA.2012-0118}{{\em
  Z.Naturforsch.} {\bfseries A68} (2013) 178--209},
\href{http://arxiv.org/abs/1207.0611}{{\ttfamily arXiv:1207.0611 [hep-th]}}.

\bibitem{Hatsuda:2013yua}
Y.~Hatsuda, M.~Honda, S.~Moriyama, and K.~Okuyama, ``{ABJM Wilson Loops in
  Arbitrary Representations},''
  \href{http://dx.doi.org/10.1007/JHEP10(2013)168}{{\em JHEP} {\bfseries 1310}
  (2013) 168},
\href{http://arxiv.org/abs/1306.4297}{{\ttfamily arXiv:1306.4297 [hep-th]}}.

\bibitem{Ouyang:2015hta}
H.~Ouyang, J.-B. Wu, and J.-j. Zhang, ``{Exact results for Wilson loops in
  orbifold ABJM theory},''
\href{http://arxiv.org/abs/1507.00442}{{\ttfamily arXiv:1507.00442 [hep-th]}}.

\bibitem{Imamura:2008dt}
Y.~Imamura and K.~Kimura, ``{N=4 Chern-Simons theories with auxiliary vector
  multiplets},'' \href{http://dx.doi.org/10.1088/1126-6708/2008/10/040}{{\em
  JHEP} {\bfseries 0810} (2008) 040},
\href{http://arxiv.org/abs/0807.2144}{{\ttfamily arXiv:0807.2144 [hep-th]}}.

\bibitem{Chen:2012at}
F.-M. Chen and Y.-S. Wu, ``{Construction of New D=3, N=4 Quiver Gauge
  Theories},'' \href{http://dx.doi.org/10.1007/JHEP02(2013)016}{{\em JHEP}
  {\bfseries 1302} (2013) 016},
\href{http://arxiv.org/abs/1210.1169}{{\ttfamily arXiv:1210.1169 [hep-th]}}.

\bibitem{Chen:2012bt}
F.-M. Chen and Y.-S. Wu, ``{Fusion of Superalgebras and D=3, N=4 Quiver Gauge
  Theories},''
\href{http://arxiv.org/abs/1212.6650}{{\ttfamily arXiv:1212.6650}}.

\bibitem{Jafferis:2008qz}
D.~L. Jafferis and A.~Tomasiello, ``{A Simple class of N=3 gauge/gravity
  duals},'' \href{http://dx.doi.org/10.1088/1126-6708/2008/10/101}{{\em JHEP}
  {\bfseries 0810} (2008) 101},
\href{http://arxiv.org/abs/0808.0864}{{\ttfamily arXiv:0808.0864 [hep-th]}}.

\bibitem{Hohenegger:2009as}
S.~Hohenegger and I.~Kirsch, ``{A Note on the holography of Chern-Simons matter
  theories with flavour},''
  \href{http://dx.doi.org/10.1088/1126-6708/2009/04/129}{{\em JHEP} {\bfseries
  0904} (2009) 129},
\href{http://arxiv.org/abs/0903.1730}{{\ttfamily arXiv:0903.1730 [hep-th]}}.

\bibitem{Gaiotto:2009tk}
D.~Gaiotto and D.~L. Jafferis, ``{Notes on adding D6 branes wrapping RP**3 in
  AdS(4) x CP**3},'' \href{http://dx.doi.org/10.1007/JHEP11(2012)015}{{\em
  JHEP} {\bfseries 1211} (2012) 015},
\href{http://arxiv.org/abs/0903.2175}{{\ttfamily arXiv:0903.2175 [hep-th]}}.

\bibitem{Hikida:2009tp}
Y.~Hikida, W.~Li, and T.~Takayanagi, ``{ABJM with Flavors and FQHE},''
  \href{http://dx.doi.org/10.1088/1126-6708/2009/07/065}{{\em JHEP} {\bfseries
  0907} (2009) 065},
\href{http://arxiv.org/abs/0903.2194}{{\ttfamily arXiv:0903.2194 [hep-th]}}.

\bibitem{Gaiotto:2008cg}
D.~Gaiotto, S.~Giombi, and X.~Yin, ``{Spin Chains in N=6 Superconformal
  Chern-Simons-Matter Theory},''
  \href{http://dx.doi.org/10.1088/1126-6708/2009/04/066}{{\em JHEP} {\bfseries
  0904} (2009) 066},
\href{http://arxiv.org/abs/0806.4589}{{\ttfamily arXiv:0806.4589 [hep-th]}}.

\bibitem{Terashima:2008sy}
S.~Terashima, ``{On M5-branes in N=6 Membrane Action},''
  \href{http://dx.doi.org/10.1088/1126-6708/2008/08/080}{{\em JHEP} {\bfseries
  0808} (2008) 080},
\href{http://arxiv.org/abs/0807.0197}{{\ttfamily arXiv:0807.0197 [hep-th]}}.

\bibitem{Bandres:2008ry}
M.~A. Bandres, A.~E. Lipstein, and J.~H. Schwarz, ``{Studies of the ABJM Theory
  in a Formulation with Manifest SU(4) R-Symmetry},''
  \href{http://dx.doi.org/10.1088/1126-6708/2008/09/027}{{\em JHEP} {\bfseries
  0809} (2008) 027},
\href{http://arxiv.org/abs/0807.0880}{{\ttfamily arXiv:0807.0880 [hep-th]}}.

\bibitem{Peskin:1995ev}
M.~E. Peskin and D.~V. Schroeder, {\em {An Introduction to quantum field
  theory}}.
\newblock Westview Press, Boulder, Colorado, USA,
1995.
\newblock

\end{thebibliography}

\providecommand{\href}[2]{#2}\begingroup\raggedright\endgroup

\end{document}